\documentclass[acmart,,natbib=true,anonymous=false]{acmart}
\usepackage{subfiles}
\usepackage{longtable}
\usepackage{xcolor, soul}
\usepackage{multirow}
\usepackage[title]{appendix}
\usepackage{subfigure}
\usepackage[ruled,vlined]{algorithm2e}

\AtBeginDocument{%
  \providecommand\BibTeX{{%
    \normalfont B\kern-0.5em{\scshape i\kern-0.25em b}\kern-0.8em\TeX}}}

\setcopyright{acmcopyright}
\copyrightyear{2022}
\acmYear{2022}
\acmDOI{}

\acmConference[CARS@RecSys'22]{Workshop on Context-Aware Recommender Systems}{Seattle, WA, USA}
\begin{document}
\definecolor{holden-color}{rgb}{0.2, 0.65, 0.99}
\newcommand{\holden}[1]{\textcolor{holden-color}{$_{Holden}$[#1]}}

\newcommand{\kmy}[1]{\textcolor{purple}{$_{min}$[#1]}}
\newcommand{\yd}[1]{\textcolor{pink}{$_{yiding}$[#1]}}
\newcommand{\plm}[1]{\textcolor{red}{$_{plm}$[#1]}}
\definecolor{yisong-color}{rgb}{0.2, 0.65, 0.99}
\newcommand{\yisong}[1]{\textcolor{yisong-color}{$_{Yis}$[#1]}}
% \newcommand{\yisong}[1]{}
% Min5: suggest new title
% Min5: Prerequisite Driven Recommendation
\title{Modeling and Leveraging Prerequisite Context in Recommendation}

% \title{Knowledge-driven Recommender: Do Not Forget Prerequisite Impact in Knowledge Linkage}
% \title{Joint Learning of Prerequisite Relation and User Behavior in Recommender}

% \author{Anonymous}
\author{Hengchang Hu}
\email{hengchang.hu@u.nus.edu}
\affiliation{%
  \institution{National University of Singapore}
  \streetaddress{AS6}
  \country{Singapore}
}
\author{Liangming Pan}
\affiliation{%
  \institution{National University of Singapore}
  \country{Singapore}
}
\author{Yiding Ran}
\affiliation{%
  \institution{National University of Singapore}
  \country{Singapore}
}
\author{Min-Yen Kan}
\email{kanmy@comp.nus.edu.sg}
\affiliation{%
  \institution{National University of Singapore}
  \country{Singapore}
}

\renewcommand{\shortauthors}{Hengchang Hu, et al.}

\begin{abstract}

% Music listening preferences at a given time depend on a wide range of contextual factors, such as user emotional state, location and activity at listening time, the day of the week, the time of the day, etc. It is therefore of great importance to take them into account when recommending music. However, it is very difficult to develop context-aware recommender systems that consider these factors, both because of the difficulty of detecting some of them, such as emotional state, and because of the drawbacks derived from the inclusion of many factors, such as sparsity problems in contextual pre-filtering. This work involves the proposal of a method for the detection of the user contextual state when listening to music based on the social tags of music items. The intrinsic characteristics of social tagging that allow for the description of items in multiple dimensions can be exploited to capture many contextual dimensions in the user listening sessions. The embeddings of the tags of the first items played in each session are used to represent the context of that session. Recommendations are then generated based on both user preferences and the similarity of the items computed from tag embeddings. Social tags have been used extensively in many recommender systems, however, to our knowledge, they have been hardly used to dynamically infer contextual states.

% 除了时间地点等，之前积累的用户信息（user profile）进行推荐
%%%% Version 2nd %%%%%%%
% Summary
% are defined as necessary preconditions
Prerequisites can play a crucial role in users’ decision-making 
% MinCR: not sure this clause helps.  What is it trying to do? Fixed differently.
% \holden{while providing background knowledge}, 
yet recommendation systems have not fully utilized such contextual background knowledge. Traditional recommendation systems (RS) mostly enrich user--item interactions where the context consists of static user profiles and item descriptions, ignoring the contextual logic and constraints that underlie them. 
% MinCR: Rewritten. You have simple spelling and POS errors, please try harder not to introduce new errors with your fixes.
% \holden{For example, 
% the recommendation of an item to a user may be only in the constrain that the user has interacted with % another item, due to the preconditinal knowledge he/she has obtained.}
For example, a RS may recommend an item on the condition that the user has interacted with another item as its prerequisite. 
Modeling prerequisite context from conceptual side information can overcome this weakness. 
We propose Prerequisite Driven Recommendation (PDR), a generic context-aware framework where prerequisite context is explicitly modeled to facilitate recommendation. We first design a Prerequisite Knowledge Linking (PKL) algorithm, to curate datasets facilitating PDR research. Employing it, we build a 75k+ high-quality prerequisite concept dataset which spans three domain.
We then contribute PDRS, a neural instantiation of PDR. By jointly optimizing both the prerequisite learning and recommendation tasks through multi-layer perceptrons, we find PDRS consistently outperforms baseline models in all three domains, by an average margin of 7.41\%. 
Importantly, PDRS performs especially well in cold-start scenarios
with improvements of up to 17.65\%. 
% We further conduct an in-depth study of our induced prerequisite graphs,
% finding intrinsically-motivated induced prerequisite chains. 
% Our extrinsic validation of PDRS shows that the fine-grained modeling of prerequisite knowledge from both the user and item perspectives result in varying efficacy across domains under cold start.  

\end{abstract}





\maketitle

% \subfile{sec/intro.tex}
\section{Introduction}
\begin{figure}[t!]
\setlength{\abovecaptionskip}{-0.01cm}
\setlength{\belowcaptionskip}{-0.6cm}
\centering
\includegraphics[width=0.88\linewidth]{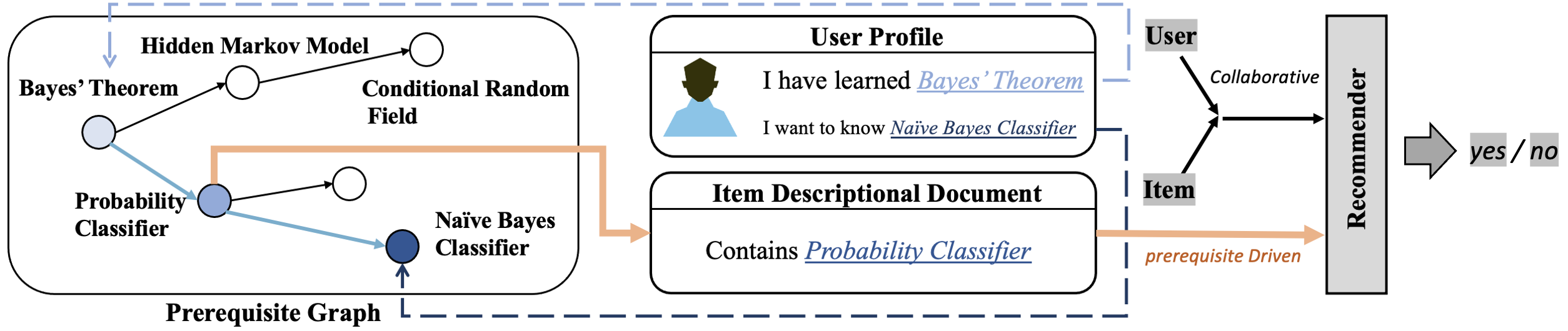}
\caption{An illustration of prerequisite driven recommendation: in which a recommender (right) incorporates prerequisite knowledge (left), distilled from both user and item prerequisite context.
Dashed edges link users to the concepts they have mastered as prior concepts and the desired concepts to acquire as target knowledge. }
\label{fig:example}
\end{figure}

{\it Prerequisites} are defined as the necessary contexts that enable downstream activity or state in human cognitive processes \cite{laurence1999concepts}. In certain domains --- especially education \cite{ohland2004identifying,vuong2011method,agrawal2016toward} --- such requisites are an important consideration that constrains item selection.
% Review
Context-aware recommendation systems have integrated the use of collaborative filtering with auxiliary metadata about users' current background or state, such as time sand location \cite{sun2019research, livne2019deep}.
% | Generally, contextual information is incorporated to facilitate recommending items to users under certain circumstances, such as the temporal context.
% | The dynamicity of context would influence user decision under certain circumstances.
% | In this work, we aim to point out another specific context -- prerequisite context in the knowledge background, where different people at different stage handle different level of knowledge.
% Recommendation systems (RS) have integrated the use of collaborative filtering with explicit auxiliary metadata about users and items \cite{sun2019research} (termed {\it context}).
% Gap
However, the role of prerequisite context (represented in the form of concepts \cite{laurence1999concepts} describing items) has been neglected in recommendation, where such dependent information is crucial for modeling users' interests.
% \holden{In our work, we term the knowledge background of a user as a certain context as well, and figure out how to leverage it through the prerequisite graphs.}
Take the educational domain example in Figure~\ref{fig:example}:
The item's key concept \textit{Probability Classifier} and user's prior knowledge are both prerequisite contexts for recommendation.
By leveraging the implicit prerequisite relationships between them (represented in the form of a prerequisite graph), we can achieve a comprehensive and explainable recommender that connects the user prerequisite context with item prerequisite context.
In our example, \textit{Probability Classifier} is beyond the knowledge of what the user already knows ({\it Bayes' Theorem}) but on the path towards to the user's desired target knowledge (\textit{Naive Bayes Classifier}), thus deserving a higher recommendation probability.
Comparatively, items containing the concept of {\it Conditional Random Field} --- although related to {\it Bayes' Theorem} --- would be poor choices since the user lacks the prerequisite prior knowledge of {\it Hidden Markov Model}. 
% Purpose
Given the importance of user and item prerequisite context, we explore the possibility of modeling and leveraging them in recommendation.

Though previous work \cite{hidasi2015session} focused on the sequential modeling between items, there is virtually no work investigating the next-step decision in light of users' conceptually-mastered knowledge. We fill this gap by capturing
% We capture 
the user and item prerequisite contexts through the compilation of a prerequisite graph, and treating the user's knowledge as static context.
Specifically, both the set of {\it prior concepts} that a user has already mastered, and the set of {\it target concepts} the user aims to acquire, directly influence the sequence of items to recommend.
% For example, in course recommendation, it is more reasonable to recommend a course when the user has acquired most of the prerequisite concepts it requires, and where the prospective course also contains the knowledge that the user desires to learn. 
Based on this, we make a key observation that prerequisite driven recommendation requires two subtasks: (1) prerequisite modeling at the concept level, 
% yiding: is 'concept prerequisite' equivalent to 'prerequisite context'
and (2) user modeling that identifies the prior and target concepts at the user--item level. 
% yiding: why is the seocond subtask at the user--item level?
More concretely, concept prerequisite modeling is the identification of prerequisite links among concepts; {\it cf} Fig.~\ref{fig:example}, prerequisite edges link the introductory concept {\it Probability Classifier} to the more advanced {\it Na\"{\i}ve Bayes Classifier}. Prior and target concept identification is thus the the process of inferring the state of knowledge for each user, with respect to the inventory of concepts in the prerequisite graph. 

% Method
To demonstrate the effectiveness of leveraging prerequisite context for recommendation, we propose a Prerequisite-Driven Recommendation System (PDRS) embodying this  formalism. 
%coupling it to traditional collaborative mechanisms. 
The key challenge here is how to accurately acquire user- and item- forms of prerequisite context for use in recommendation. 
% inferred from the user--item interaction matrix
However, there is an important shortcoming.  Explicit prerequisite knowledge is often sparse, requiring laborious effort to compile.  It is often also brittle, as items and their relationships with underlying contextual concepts can evolve over time.  
Assuming manually-labeled prerequisites \cite{vuong2011method,talukdar2012crowdsourced} is often unrealistic due to the heavy cost of human annotation.  An automatic means of inferring prerequisites is called for.
% Holden: @Min. Need double-check below.
% MinCR: proofread looks good.
We address this in two parts by contributing a) an automatic key knowledge concept extractor from item descriptive text, and b) a prerequisite relation constructor for concept pairs by inferring prerequisite weights from both internal and external domain features.
Our encoding components finds a suitable representation of a user in terms of prior and target knowledge, leveraging pretrained language models. We then integrate such prerequisite context into the recommendation process by joint training of both the recommendation and prerequisite knowledge learning tasks.
% as well as users' prior and target knowledge. 
% Our Prerequisite Context Modeling component constructs users' prior and target knowledge ...
% The prerequisite context leverage the knowledge from side prerequisite knowledge (in the form of topologically-sorted prerequisite graph) 

We evaluate our PDRS system on three datasets representing different domains. We find that PDRS achieves performance gains in recommendation not only in domains where prerequisites exist as hard constraints --- such as (educational) course recommendation --- but also in domains where prerequisites are soft, personal preferences, as in movie and book recommendation. Importantly, we also show that such model makes an especially strong impact in sparse data cold-start scenarios, a pervasive problem in RS. 

% We summarise our contributions as follows: 
% \begin{itemize}
%     \item To the best of our knowledge, we are the first to explore the use of prerequisite context for recommendation.
%     % \item We formalise the framework of prerequisite-driven recommendation consisting three components: 1) the users' prerequisite context and 2) the items' prerequisite context, and 3) the prerequisite graph;
%     \item We instantiate our formalism in the form of a Prerequisite Driven Recommendation System (PDRS; \S~4) embodied as a modern neural architecture, which adopts joint training to optimise the model for the twin objectives of knowledge linking prediction and recommendation;
%     \item We demonstrate that prerequisite context is a functional booster solving cold-start problem, and can benefit recommenders universally through our extensive experiments on our Course, Movie, and Book datasets (\S~5).
% \end{itemize}

\section{Problem Formulation}
\label{sec:problem_formulation}

% PDR has the recommendation task as a subgoal: to recommend an item $v$ for user $u$ as estimating $u$'s preference by interaction history stored in $\mathbf{Y}$, where $y_{uv}=1$ means $u$ has interacted with $v$.

% \vspace{0.1cm}

\noindent \textit{Definition 1. Prerequisite Context.}
The context of each user $u$ can be seen as a personal concept repository consisting of a set of mastered prior concepts and a set of target concepts to be acquired, represented as sets $\{\mathcal{C}_u^p, \mathcal{C}_u^t\} \subseteq \mathcal{C}$, respectively. 
Items then contain these concepts: each item $v$'s context are denoted as $\mathcal{C}_v^i$.  Importantly, items manifest concepts in a latent, implicit manner, such that the item concept inventory must be inferred. 
% where ideal items to be recommended should include concepts bridge the prior to the target. 

% Through the prerequisite graph, they can be linked to the knowledge concepts contained in each item $v$'s context, denoted as $\mathcal{K}_v^i$. The linkages include prerequisite strength from user's prior knowledge and the target knowledge, where ideal items to be recommended bridge the prior to the target. 

\vspace{0.1cm}
\noindent \textit{Definition 2. Prerequisite Graph.} We represent prerequisite context as a graph $\mathcal{G}$, having concepts $\mathcal{C}$ as nodes, and prerequisite relations $\mathcal{R}$ among them as edges, where $r \in \mathcal{R}$ represents the confidence towards prerequisite relation linking knowledge $c_p$ to knowledge $c_q$. % ($c_p, c_q \in \mathcal{C}$).  
$\mathcal{G}$ can thus be represented as a series of edge tuples;
for example, \textit{(logic, 0.99, python)} means that \textit{logic} is prior knowledge required for \textit{python}
with a confidence score of 0.99.

% \begin{table}[t]
%     \centering
%     \small
%     \begin{tabular}{cc || cc}
%     % \hline
%     \toprule
%     {\bf Notation} & {\bf Explanation} & {\bf Notation} & {\bf Explanation} \\
%     % \hline
%     % \hline
%     \midrule
%     $\mathcal{U} = \{u_1, \cdots, u_n\} $ & Set of users  &  
%     $\mathcal{R} = \{r_1, \cdots\} $ & Set of relations over knowledge pairs\\
%     $\mathcal{V} = \{v_1, \cdots, v_m\} $ & Set of items &
%     $\mathcal{G} = (\mathcal{K}, \mathcal{R}) $ & Prerequisite graph for storing dependencies \\
%     $\mathbf{Y} $ & User--item interaction matrix &
%     $\mathcal{K}_u^p, \mathcal{K}_u^t \subseteq \mathcal{K} $ & Prior and Target knowledge set for user $u$\\
%     $\mathcal{H}_u \subseteq \mathcal{V}$ & $u$'s historical interaction sequence &
%     $\mathcal{K}_v^i \subseteq \mathcal{K} $ & Knowledge contained in or related to item $v$\\
%     $\mathcal{K} = \{k_1, \cdots, k_h\} $ & Set of all knowledge concepts &
%     $\mathcal{D}_v $ & Document of item $v$\\
    
%     \bottomrule
%     \end{tabular}
%     \caption{Notation used for prerequisite representations.} 
%     \label{tab:symbol} 
%     \vspace{-7mm}
% \end{table}

% \vspace{0.1cm}

% \noindent \textit{Definition 3. Prerequisite-driven recommendation (PDR)}. 
The task of Prerequisite-Driven Recommendation (PDR) aims to learn the latent factors not only from user--item interactions $\mathbf{Y}$ (where $y_{uv}=1$ means $u$ has interacted with $v$), but also from prerequisite context.
We can view PDR as combining the knowledge linkage prediction task $g$ and context-aware recommendation prediction task $f$. 
Specifically, it can be formalized as: 1) inferring latent prerequisites $\hat{r}_{c_ic_j}=g(c_i,c_j|\Phi,\mathcal{G})$, where $\hat{r}_{c_ic_j}$ represents the predicted prerequisite confidence from concept $c_i$ to $c_j$; and 2) prerequisite-driven recommendation $\hat{y}_{uv}=f(u,v,\{c|c\in (\mathcal{C}_u^p \cup \mathcal{C}_u^t \cup \mathcal{C}_v^i)\} |\Theta,\Phi, \mathbf{Y},\mathcal{G})$.  Here, $\Phi$ and $\Theta$ are the parameters for encoding knowledge concepts and users/items. 

% \vspace{0.1cm}

\begin{figure}[t] 
    \setlength{\abovecaptionskip}{-0cm}
    \setlength{\belowcaptionskip}{-0.5cm}
	\subfigtopskip=2pt 
	\subfigbottomskip=2pt 
	\subfigcapskip=-5pt
	\subfigure[Prerequisite Knowledge Inference]{
	    \centering
		\label{level.sub.1}
		\includegraphics[width=0.50\linewidth]{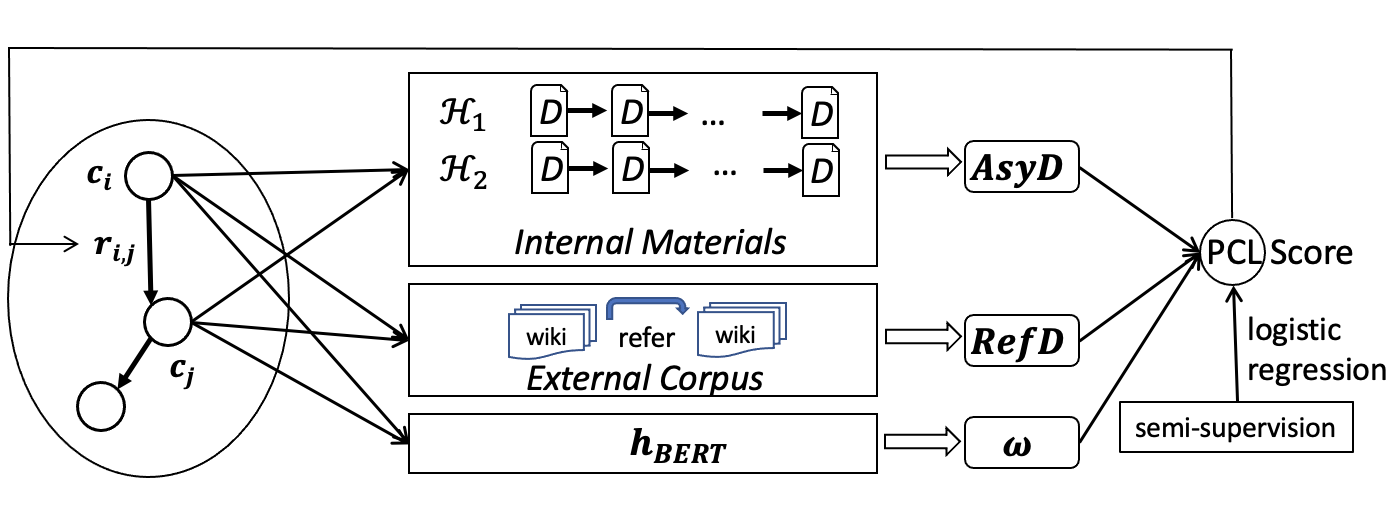}}
% 	\quad 
	\subfigure[Prerequisite Context Modeling in Recommendation]{
		\label{level.sub.2}
		\includegraphics[width=0.48\linewidth]{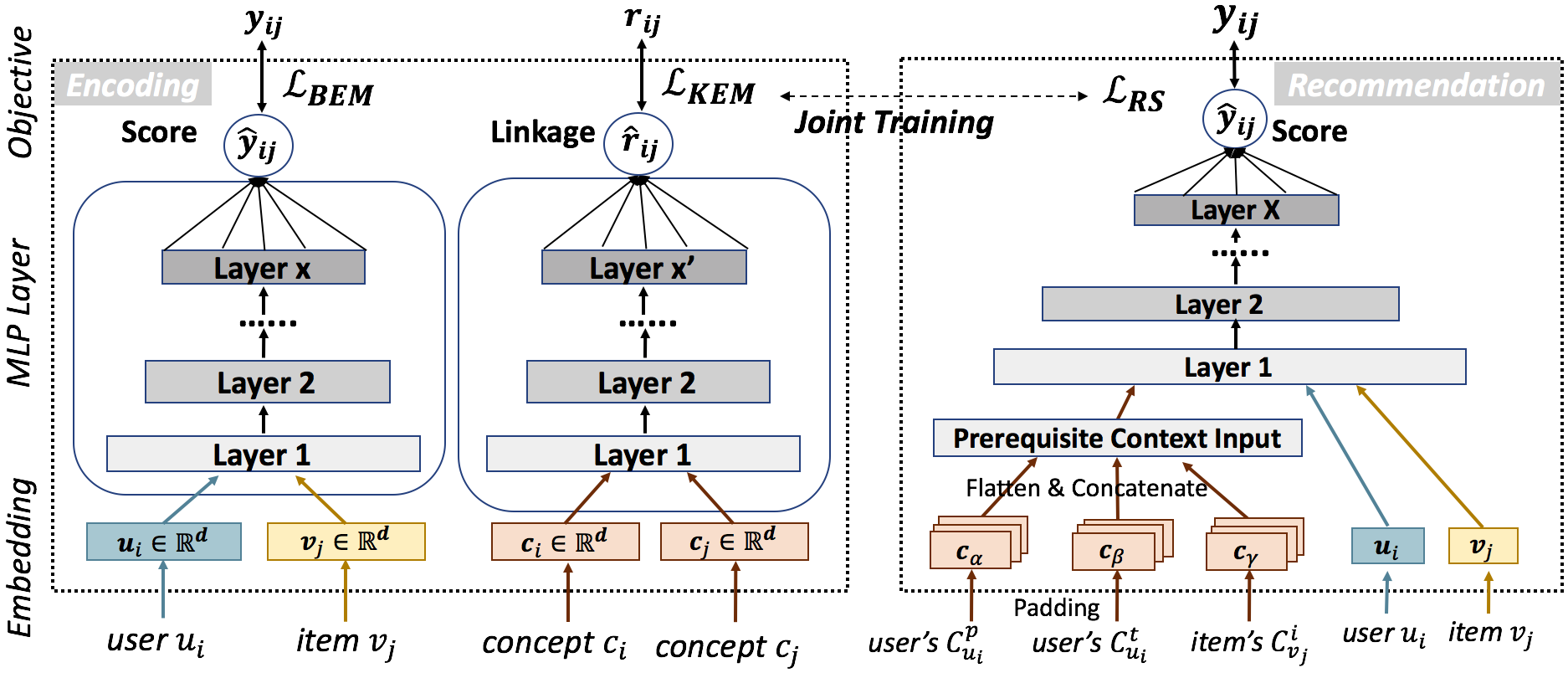}}
	\caption{Our instantiated neural PDRS framework. It consists components of a) prerequisite constructor,  b [left]) twin encoders pretrain embedding of users, items (BEM), and prerequisites (KEM), and b [right]) a fine-tuned neural recommendation system.}
	\label{fig:framework}
\end{figure}

\section{PDRS: Prerequisite-Driven RS}
%We present a simple yet effective instantiation of PDR, which guides the recommendation task with the encoded prerequisite context.
%To imbue recommendation to capture prerequisites knowledge in a PDR task, 
Our instantiated framework (PDRS) is depicted in Figure~\ref{fig:framework}.  It consists of three components (\S 3.1, 3.2~and~3.3):

\subsection{Prerequisite Knowledge Inference}
To leverage prerequisite context, we first need to build a \textit{prerequisite graph} with concept-level prerequisite links.
We decompose this process into two subtasks: 
extracting concepts from item descriptions, 
and inferring prerequisite relation between concepts from the ordered item documents and general knowledge (topological relations in Wikipedia).

% \vspace{0.12cm}
% \noindent 
\textbf{Concept Extraction.}
\label{sec:data_construct}
% Our concept extraction procedure identifies and links concepts into a graphical network.  
To extract key concepts from item documents (item description title, and item description content) as prerequisite context,
we extend prior work \cite{pan2017course} on graph propagation as a three-step subprocess.
First, \textit{seed concepts} are extracted from item $v$'s document titles using TextRank \cite{mihalcea2004textrank} (with an empirically tuned threshold of 50\%).
% Min: this clause is more confusing than helpful.  Omit.
% which indicate $v$'s central topics. 
Next, \textit{candidate concepts} are gathered from $v$'s document content, identifying all phrases that match the part-of-speech tag pattern for a noun phrase: $((A|N)+|(A|N)^{*}(NP)?(A|N)^{*})N$ \cite{justeson1995technical} (here, $A$, $N$, and $P$ indicate adjectives, nouns, and prepositions). 
% Min: what is seed-candidate? seed-relevant? Confusing.  I don't know what these qualifiers mean.  These lines 46--49 need re-writing and careful editing that doesn't introduce more difficulties. 
% Holden: added description for clarification below.
Lastly, we construct a fully-connected graph to include both seed concepts and candidate concepts, and expand the seed concept set to cover its relevant concepts. We implement this by iterative propagation, where concepts' confidence scores
(where seed concepts are initially weighted with unit confidence)
are propagated to their neighbors\footnote{For computational efficiency, we only propagate when edge scores are above a tunable parameter $\lambda$.} regarding their semantic relatedness, as measured by cosine similarity $\omega(c_{i},c_{j})=cosine(h_{BERT}(c_{i}),h_{BERT}(c_{j}))$ of the textual meaning $h$ from BERT \cite{devlin2018bert}. 

% \vspace{0.12cm}
% \noindent 
\textbf{Inferring Prerequisites.} Knowledge concepts are then associated by prerequisite relations. 
As shown in Fig~\ref{fig:framework}(a), we obtain the final linking scores by considering three features: 
i) asymmetric sequential interaction distance, 
ii) reference distance from general domain,
and iii) semantic relatedness (as defined above).
We explain the former two features:

% \vspace{1.2mm}
\textit{(i) Asymmetric Sequential Interaction Distance.}
We observe that concepts covered in subsequently-consumed
items are often dependent on those in previous ones; i.e., they are prerequisites. 
% Min: BUG choose a concept in your Fig. 1.  Max Likelihood is no longer there. Holden: Fixed
For example, courses that relate to \textit{Bayes' Theorem} are more likely to appear in a user's interaction history \textit{before} those related to \textit{Hidden Markov Model}.
We introduce Asymmetric Sequential Interaction Distance (AsyD) to model this, which captures the distributional pattern of knowledge concept pairs.
Let's say $P(c_i,c_j) = \sum_{u} \sum_{(v_{p}, v_{l}) \in \mathcal{P}_u} tf(c_i, v_{p}) tf(c_j, v_{l})$ indicates the probability of $c_i$ preceding $c_j$ in interaction histories, where $\mathcal{P}_u = \{(v_a,v_b) | v_a, v_b \in \mathcal{H}_u; ts_u(v_a)<ts_u(v_b)\}$ indicates item pairs ordered by timestamp $ts$ in user historical interaction sequence $\mathcal{H}_u$ and $tf(c,v)$ denotes the term frequency of $c$ in item $v$'s documents (e.g., title and description). 
Specifically, if $c_i$ is a prerequisite of $c_j$, the probability $P(c_i,c_j)$ should be greater than the 
% Min: converse.  Please make sure you know these three terms https://www.varsitytutors.com/hotmath/hotmath_help/topics/converse-inverse-contrapositive
converse association $P(c_j,c_i)$.
Thus, our defined Asymmetric Sequential Interaction Distance is a normalized probability: 
$AsyD\left(c_i, c_j\right) = \sigma \{P(c_i,c_j) / P(c_j,c_i) -1\}$.
% \begin{small}
% \begin{equation}
% \setlength{\abovedisplayskip}{1pt}
% \setlength{\belowdisplayskip}{1pt}
%     AsyD\left(k_i, k_j\right) = \sigma \left(\frac{\sum_{u \in \mathcal{U}} \sum_{\left(v_{p}, v_{l}\right) \in \mathcal{P}_u} tf\left(k_i, v_{p}\right) tf\left(k_j, v_{l}\right)}{\sum_{u \in \mathcal{U}} \sum_{\left(v_{p}, v_{t}\right) \in \mathcal{P}_u} tf\left(k_j, v_{p}\right) tf\left(k_i, v_{l}\right)} -1 \right)
% \end{equation}
% \end{small}
% \noindent where $\mathcal{P}_u = \{(v_a,v_b) | v_a, v_b \in \mathcal{H}_u; ts_u(v_a)<ts_u(v_b)\}$ indicates sequential item pairs in user history and $tf(k,v)$ denotes the term frequency of $k$ in item $v$. 
% ($\mathcal{H}_u \subseteq \mathcal{V}$ here represents $u$'s historical interaction sequence.)
% $AsyD\left(k_i, k_j\right)$ is a normalized probability of $k_i$ preceding $k_j$ in interaction histories, where each occurence event is weighted by the concept's frequency in the item's documents (e.g., title and description).  
When $c_i$ and $c_j$ have arbitrary consumption order --- \textit{i.e.}, when they are independent of each other's distribution --- then $AsyD$ = $0.5$.

\textit{(ii) Wikipedia Reference Distance.}
Textual evidence of prerequisites in domain documents can be sparse, resulting in noisy learned prerequisites.  
Contextual knowledge from general information sources, such as Wikipedia, can aid the identification of prerequisite relations by providing supplemental evidence.  
We observe that related general information that refer to domain concepts can also provide statistical evidence of prerequisites. 
To capture this, we propose a domain-adaptive Reference Distance ($RefD$) which builds on Liang's work \cite{liang2015measuring} on Wikipedia. 
% MinCR => Hengchang: can you give a hypothetical example of $t$ related to $c$ in Fig 1?
%% >> 
For example, the in-domain concept $c$ -- \textit{"probability classifier"} may not occur in Wikipedia, we measure its reference imbalance through its related concepts $t$ -- \textit{"bayes classifier"} in Wiki corpus.
Specifically, when a related concept $t$ is present in Wikipedia that refers to concept $c$, we capture its reference ratio, similar to $AsyD$, by calculating its normalized distance score.
% yiding: how do you identify their related Wikipedia concepts? Holden: add words to explain it. Holden: I don't want to add words to introduce more confuse. it is explained later below.
% The relatedness between $k$ and $t$ is measured by semantic similarity $\omega$.
% For a concept $k$, we consider its related concepts in Wikipedia as $t$, where the similarity is measured by the aforementioned semantic relatedness $\omega(t,k)$: 
$RefD(c_i, c_j)$ measures the imbalance between each pair of in-domain concepts $c_i$ and $c_j$ via its associated reference imbalance $\tau$ to general concepts $t$. This process transforms the concept linking task to one of locating related concepts in Wikipedia. 
The reference imbalance $\tau=1$ if and only if $t_{i}$ refers to $t_{j}$ in any Wikipedia article but where $t_{j}$ never refers to $t_{i}$. As such, $\tau$ analogously ranges $[-1,+1]$. Thus, $RefD$ can smoothly incorporate information from general sources ($\tau$), as well as in-domain sources (represented by $AsyD$); i.e., 
% Formally, $RefD(c_i, c_j)=\mathcal{N} \cdot \sum_{i,j} \tau(t_i,t_j) \cdot S_{i,j}$, where $S_{i,j}= \omega(t_i,c_i) \omega(t_j,c_j)$ is used to represent how closely $c_i$ and $c_j$ are linked to their related Wiki concepts, and $\mathcal{N}=(\sum_{i,j} S_{i,j})^{-1}$ is the normalization term.
$RefD(c_i, c_j)=\sum_{i,j} \tau(t_i,t_j) \cdot S_{i,j}/\sum_{i,j} S_{i,j}$, where 
$S_{i,j}= \omega(t_i,c_i) \omega(t_j,c_j)$ is used to represent how closely $c_i$ and $c_j$ are linked to their related Wiki concepts, and the denominator allows $RefD$ to also be interpreted as a normalized probability. 
% MinCR: why in parens?  Demote to footnote?
% HC: removed the parens, and re-contruct the sentence
% $S_{i,j}= \omega(t_i,c_i) \omega(t_j,c_j)$ is used to represent how closely $c_i$ and $c_j$ are linked to their related Wiki concepts.

% $RefD(k_i, k_j)$ indicates the imbalance between $k_i$ and $k_j$ by measuring the reference imbalance between $t_p$ and $t_l$.
%% yiding: what does 'reference imbalance' mean?
%% Min: Agree with Yiding.  This is not clear.  Math is hurting the understanding of what you are doing.  This section also needs careful re-editing.  I don't understand.
% Holden: re-edited
% Formally, $RefD(k_i, k_j)= \{\sum_{t_p,t_l} [\tau(t_p,t_l)-\tau(t_p,t_l)] S_{p,l}(k_i,k_j)\} \cdot \{\sum_{t_p,t_l}S_{p,l}(k_i,k_j)\}^{-1}$, where the reference indicator $\tau(t_p,t_l)$ is equal to 1 if $t_{p}$ refers to $t_{l}$ in an article, and 0 otherwise.
% $S_{p,l}(k_i,k_j) = \omega(t_p,k_i) \cdot \omega(t_l,k_j)$ .
% Thus the second part can be termed as the normalization.

% \vspace{0.12cm}
\textit{Prerequisite Learning.}
To obtain the final prerequisite knowledge linkage scores $PKL(c_i,c_j)$, we train logistic regression over seed annotated samples ($n = 300$ per dataset), as shown on the right part of Fig.~\ref{fig:framework}a. The regression is learned by adjusting the contribution weights for the three features $\omega, AsyD and RefD$ on manually-labeled concept pairs.  We then run the regression to yield output for all knowledge concept pairs. This process reduces noise from the parameters, providing more accurate PKL scores for downstream recommendation (Fig.~\ref{fig:framework}b).
% Our constructed prerequisite data are available at https://github.com/HoldenHu/PDRS/.

\subsection{User/item and Prerequisite Context Encoding}
\label{sec:module2}

The Encoding modules take the sparse input representations of users, items and concepts identified by PKL and encode them into dense representations
% for better performing similarity distances \cite{mikolov2013efficient}.
% yiding: what's 'better performing similarity distances'? easy computation of similarity measures?
(Fig.~\ref{fig:framework}b [left]), employing a multi-layer perceptron (MLP) for both \textit{Knowledge Encoding Module (KEM)} and \textit{Behavior Encoding Module (BEM)}.

% Holden: @Min to recheck this expression below.
% yiding: the following paragraph is a bit confusing 
KEM learns the knowledge concept embedding by training pairs of $(c_i,c_j)$ to approximate their previously-assigned prerequisite PKL scores. 
% Min: how does the math help here?  It is confusing
% Formally, it optimizes the parameters $\operatorname{argmin}_{\Phi} \sum_{(i,j)} \mathcal{L}_{KEM}(r_{i,j}, \hat{r}_{i,j})$, where $r_{i,j}$ is expected prerequisite weight $PKL(k_i,k_j)$, and $\hat{r}_{i,j}$ is the estimated output from a $x$-layer $MLP_{x}({k_i}, {k_j}, \Phi)$ which interacts the factors from concatenated embedding of $k_i$ and $k_j$.
Our target in this step is to obtain every concept $c$'s embedding $\mathbf{c} \in \mathbf{\mathbb{R}^{1 \times d}}$ by tuning the learnable parameters $\Phi$
to achieve the minimized $\mathcal{L}_{KEM}(r_{i,j}, \hat{r}_{i,j})$, where $r_{i,j}$ is expected prerequisite weight $PKL(k_i,k_j)$, and $\hat{r}_{i,j}$ is the estimated output from a 
% MinCR: I don't think the x notation helps to make anything more clear.  Dropping.  Please check.
% $x$-layer MLP.
MLP.
% \vspace{-0.1cm}
BEM works similarly, learning user/item embeddings by minimizing the difference $\mathcal{L}_{BEM}$ between predicted 
% MinCR: why x'?  Obsolete anyways.
% $\hat{y}_{i,j} = MLP_{x'}(u_i, v_j, \Theta)$ 
$\hat{y}_{i,j} = MLP(u_i, v_j, \Theta)$ 
and the truth $y_{i,j}$. 
% Min: again the math doesn't help (to me). 
% Formally, it targets at learning $\operatorname{argmin}_{\Theta} \sum_{(i,j)} \mathcal{L}_{BEM}(y_{i,j}, \hat{y}_{i,j})$ projecting $u_i$, $v_j$ to their embedding $\mathbf{u}_i$, $\mathbf{v}_j$ separately.
We select a mean-squared loss for $\mathcal{L}_{KEM}$
% $ = \sum_{i,j}(\hat{r}_{i,j}-r_{i,j})^{2}$  
to best model the real-valued prerequisite scores, and a binary cross entropy loss for $\mathcal{L}_{BEM}$.
%$ = -\sum_{i,j} (y_{i,j} \times ln (\hat{y}_{i,j}) +(1-y_{i,j} \times ln(1-\hat{y}_{i,j})))$.

%%%%%%%%%%%%%%%%%%%%%%%%%%%%%%%%%%%%%%%%%%%%%%%%%%%
% Module 3
%%%%%%%%%%%%%%%%%%%%%%%%%%%%%%%%%%%%%%%%%%%%%%%%%%%
\subsection{Recommendation Module}
\label{sec:module3}
As shown in Fig.~\ref{fig:framework}b [right],
PDRS combines the embedded representation of user context (user prior knowledge concepts $C^p_u$ and target knowledge concepts $C^t_u$) and item context (item concepts $C_v^i$) to output the final recommendation prediction $\hat{y}_{ij}$. We apply the average pooling for the concept embedding, while a user/item corresponds to multiple knowledge concepts. Formally, $\hat{y}_{uv}=MLP\{\mathbf{u}, \mathbf{v}, avg(\mathbf{C}^p_u), avg(\mathbf{C}^t_u), avg(\mathbf{C}_v^i)\}$,
% user context (user's prior and target knowledge), and item context (item contained knowledge)
% We minimise difference between ground truth and prediction, using the embedded representation of the input: 3-tuples consisting of a concatenation of the average embedding of 1) user prior knowledge $K^p_u$, 2) target knowledge $K^p_t$, 3) item knowledge $K$.
% \vspace{-0.1cm}
% \paragraph{Pre-training.} 
where BEM and KEM provide the pre-trained embeddings $\mathbf{u}$, $\mathbf{v}$, and $\mathbf{c}$ for the user, item and prerequisite contexts, respectively.
% Outputs from the encoding process (KEM and BEM) initialize encoding layers $\mathbf{K}, \mathbf{V}, \mathbf{U}$ to map $k,v,u$ to their corresponding embeddings $\mathbf{k},\mathbf{v},\mathbf{u}$, as a form of pre-training.  
We further tune the embedding parameters through joint training that alternately applies the objective of $\mathcal{L}_{KEM}$ and the final recommendation $\mathcal{L}_{Rec}$.  $\mathcal{L}_{Rec}$ is also a binary cross entropy loss between model prediction $\hat{y}$ and the ground truth interaction $y$.
% When training the recommendation module, $\mathbf{K}, \mathbf{V}, \mathbf{U}$ are then fine-tuned.

% \vspace{-0.1cm}
% \paragraph{Joint Training}
% The training of the recommendation module optimizes $\mathbf{K,V,U}$ for only the recommendation task. As a side effect, this optimisation weakens the prerequisite information stored in $\mathbf{K}$, gradually learning collaborative patterns. For example, if users tend to study \textit{Python} and \textit{Photoshop} together, their learned representations will be similar in $\mathbf{K}$.

% We design PDRS to benefit primarily from the pre-trained prerequisite information.  As such. we expect the pre-trained $\mathbf{K}$ should not change much.  Freezing such pre-trained weights is possible, but shown to be suboptimal \cite{kwok1993experimental}.
% Experimental analysis of input weight freezing in constructive neural networks
%% Min5: consider showing how much this alternating training actually helps performance.  Is it an important detail? 
% We solve this by alternately training for two objectives: a mean-squared loss for prerequisite representation ($\mathcal{L}_{KLP}$; in the Encoder) and a binary cross entropy loss for final recommendation ($\mathcal{L}_{RS}$; in the Recommender):

% \vspace{-0.3cm}
% \begin{eqnarray}
% \setlength{\abovedisplayskip}{-0.1pt}
% \setlength{\belowdisplayskip}{0pt}
%     \mathcal{L}_{KLP}= &\sum_{i,j}(\hat{r}_{i,j}-r_{i,j})^{2} \\
%     \mathcal{L}_{RS}= & -\sum_{i,j} (y_{i,j} \times ln (\hat{y}_{i,j}) +(1-y_{i,j} \times ln(1-\hat{y}_{i,j}))) 
% \end{eqnarray}

We empirically observe that the joint training of losses $\mathcal{L}_{KEM}$ and $\mathcal{L}_{Rec}$ enables the model to focus on prerequisite information.  It also reduces the time complexity to a linear $O(|Y|+|R|)$ from the total quadratic complexity $O(\frac{|Y| \times |R| \times (dY+dR)}{dY \times dR})$ of alternatively executing the two objectives (where $dY$ and $dR$ represent the batch sizes of interaction and prerequisite pairs, respectively).

\section{Experiments}
\label{s:exp}

% We compare the impact of prerequisite incorporation under different recommendation scenarios, then analyze results focusing mainly on course recommendation, as it is the most knowledge-intensive scenario.

% \subsection{Experimental Settings}
% \subsubsection{Datasets.} 
To the best of our knowledge, no existing datasets are specifically designed for \textit{prerequisite context} modeling.  
Hence, we modify the existing datasets SSG-Data, MovieLens, and Amazon Books\footnote{(S) \url{https://www.skillsfuture.gov.sg/}, (M) \url{https://grouplens.org/datasets/movielens/}, and (A) \url{https://jmcauley.ucsd.edu/data/amazon/} \label{fn:web}}
which all contain textual descriptions of items.
SSG-Data (\textbf{Course}) is an exhaustive anonymized listing of life-long course-taking history of citizen participants in 9 month period in \textit{SkillsFuture Singapore (SSG)}, which is not publicly available yet. Course \textit{content} and \textit{objectives} are used as an item's documents. 
We also use the public MovieLens 100K and Amazon Books as {\bf Movie} and {\bf Book} datasets, where the crawled textual description from IMDB and TMDB\footnote{\url{https://www.imdb.com}, and \url{ https://themoviedb.org}} serve as movies' documents, and the crawled textual overview from Goodreads and Google Books\footnote{\url{https://www.goodreads.com/} and \url{https://books.google.com/}} serve as books' documents.
To align the task with (binary) recommendation, where only implicit feedback is available, we deem interactions with $ratings \ge 3$ as positive feedback in the movie and book scenarios. We provide detailed dataset statistics in Appendix~A. 
% \footnote{https://holdenhu.github.io/assets/pdf/pdrs\_appendix.pdf}

To obtain the user's state of knowledge $\mathcal{C}_u^p$, $\mathcal{C}_u^t$, 
we assume that users have mastered the knowledge contained in items that they have previously interacted with,
and take concepts from the
documents of first 30\% and the last 20\% of items they interacted with as their prior and target knowledge, respectively.
To maintain strict training and testing separation, we only use the remaining 50\% of knowledge concepts for training and testing our PDRS. 
We also follow the common practice \cite{lei2020interactive} and only retain user records with more than three interactions, to ensure each user has at least one item for evaluating recommendation performance and one item each for modeling prior and target knowledge.

% \subsection{Baselines \& Evaluation Metrics}
We assess our PDRS method against traditional recommendation models (ItemPop, ItemKNN \cite{sarwar2001item}, and NeuMF \cite{he2017neural}) and lightweight yet effective feature-incorporated models (GCMC  \cite{berg2017graph} with optimized dropout rate $0.5$, and DeepFM \cite{guo2017deepfm} with optimized dropout rate $0.2$). 
% Defaultly, we take user and item IDs as feature inputs and set optimal knowledge concept embedding dimension empirically at $64$. 
% rewriten
% Min: re-written again by Min
We also compare against two-widely accepted models of GCMC and DeepFM to validate whether our prerequisite context representation is effective
by replacing their user and item IDs feature inputs and modifying them to accept our KEM-derived $64$-dimension feature embedding (denoted as GCMC+KEM and DeepFM+KEM).
% systematically tuned the hyperparameters of all the compared models
Specifically, the feature embedding dimensions, and node dropout rates are tuned for optimal performance, set as $\{8,8,16\}$, and $0.5$, respectively. The optimal knowledge concept embedding dimension is set empirically at $64$.
We also employ two weak baselines: BPR \cite{rendle2012bpr} and SVD \cite{koren2009matrix} for a comprehensive comparison. 

% 4. \textbf{GCMC} \cite{berg2017graph} uses the encoder--decoder framework on a bipartite graph to complete the interaction matrix. 
% We take user and item IDs as feature inputs for the encoders and decode their embeddings bilinearly. The feature embedding dimensions, and node dropout rates are tuned for optimal performance, set as $\{8,8,16\}$, and $0.5$, respectively. The optimal knowledge concept embedding dimension is set empirically at $64$. GCMC can also be modified to accept our KEM pretraining, adding the corresponding prior and target knowledge embedding and the item knowledge embedding. We denote this version as \textbf{GCMC+KEM}). 

% 5. \textbf{DeepFM} \cite{guo2017deepfm}
% is a feature-based factorization model. We combine two categorical fields --- user and item IDs --- as sparse input features.  Dropout is tuned optimally to $0.2$. Like GCMC, we can hybridize DeepFM to leverage KEM as a dense input feature (denoted as \textbf{DeepFM+KEM}). 

We split our interaction data into (80\%, 10\%, 10\%) to serve as training, validation and testing, respectively.  For studying warm-start, we leave one item out per user.  For the user (item) cold-start (also called new item) scenarios, users (items) in validation/testing set do not appear in the training set. For each test case, we follow the common practice \cite{he2017neural}, ranking 100 items --- 99 negative samples and 1 positive sample. We use Hit Ratio@k (HR@k) and Normalized Discounted Cumulative Gain@k (NDCG@k) \cite{Shani2011evaluating} as top-$k$ ranking-based accuracy measures.

% \subsection{Baselines}
% We assess our PDRS method against traditional recommendation models (Models 1--3) and state-of-the-art (SOTA) deep matrix factorization models (4--5). We also use SOTA feature-incorporated models to assess PDRS's capability of exploiting side information.

% 1. \textbf{ItemPop} recommends the most popular items based on number of interactions in the training set, for all users. 

% 2. \textbf{ItemKNN} \cite{sarwar2001item} is a memory-based \cite{yu2004probabilistic} collaborative filtering method which recommends the nearest neighbors of items a user has interacted with in the training set, based on cosine similarity of the interaction vector. 

% 3. \textbf{NeuMF} \cite{he2017neural} marries general matrix factorization with the flexibility of deep learning via multi-layer perceptron (MLP; also featured in our model), concatenating their hidden states from two channels for final recommendation. 

\subsection{Main Results}
\begin{table*}[t!]
\setlength{\abovecaptionskip}{-0.05cm}
% \setlength{\belowcaptionskip}{-0cm}
% \small
\resizebox{\textwidth}{!}{%
\begin{tabular}{p{0.5cm}l|llll|llll|llll}
\hline
\multicolumn{2}{c|}{\multirow{2}{*}{Model}} & \multicolumn{4}{c|}{Course}  & \multicolumn{4}{c|}{Movie} & \multicolumn{4}{c}{Book}                                                                                 \\
\multicolumn{1}{c}{}  & \multicolumn{1}{c|}{}  & \multicolumn{1}{c}{H@2} & \multicolumn{1}{c}{H@10} & \multicolumn{1}{c}{N@2} & \multicolumn{1}{c|}{N@10} & \multicolumn{1}{c}{H@2} & \multicolumn{1}{c}{H@10} & \multicolumn{1}{c}{N@2} & \multicolumn{1}{c|}{N@10} & \multicolumn{1}{c}{H@2} & \multicolumn{1}{c}{H@10} & \multicolumn{1}{c}{N@2} & \multicolumn{1}{c}{N@10} \\ \hline

\multicolumn{1}{c|}{\multirow{7}{*}{}} &
ItemPop & 0.4541 & 0.7813 & 0.3973 & 0.5300 & 0.2504 & 0.5905 & 0.2152 & 0.3496  & 0.2305 & 0.4740 & 0.2003 & 0.2865 \\
\multicolumn{1}{c|}{} & ItemKNN & \textbf{0.7122} & 0.8076  & \textbf{0.6551} & 0.6785 & 0.2631 & 0.4613 & 0.2299 & 0.3092 & 0.0890 & 0.1178 & 0.0818 & 0.0944 \\
\multicolumn{1}{c|}{Without} & SVD & 0.5934                   & 0.8256                     & 0.5863                    & 0.6672                       & 0.2720                   & 0.6552                     & 0.2288                    & 0.3789                       & 0.2025                   & 0.4002                     & 0.1791                    & 0.2527                      \\
\multicolumn{1}{c|}{Prereq.} & BPR                                         & 0.6650                   & 0.8396                     & 0.6317                    & 0.6794                       & 0.2882                   & 0.6673                     & 0.2469                    & 0.3958                       & 0.2048                   & 0.4043                     & 0.1820                    & 0.2712                      \\
\multicolumn{1}{c|}{Knowl.} & NeuMF                                       & 0.6859                   & 0.8455                     & 0.6257                    & 0.6811                       & 0.2899                   & 0.6641                     & 0.2467                    & 0.3936                       & 0.2178                   & 0.4404                     & 0.1876                    & 0.2748                      \\
\multicolumn{1}{c|}{} & GCMC                                        & 0.4081                   & 0.7661                     & 0.3471                    & 0.4928                       & 0.2518                   & 0.5904                     & 0.2165                    & 0.3511                       & 0.2234                   & 0.4734                     & 0.2234                    & 0.2929                      \\
\multicolumn{1}{c|}{} & DeepFM                                      & 0.5329                   & 0.8154                     & 0.4630                    & 0.5799                       & 0.2695                   & 0.6457                     & 0.2341                    & 0.3821                       & 0.2236                   & 0.4431                     & 0.1915                    & 0.2751                      \\
\hline
\multicolumn{1}{c|}{With} & GCMC+KEM                                    & 0.4512                   & 0.7996                     & 0.3821                    & 0.5241                       & 0.2618                   & 0.6066                     & 0.2266                    & 0.350                       & 0.2319                   & 0.4885                     & \textbf{0.2251}           & 0.2992                      \\
\multicolumn{1}{c|}{Prereq.} & DeepFM+KEM                                  & 0.6786                   & 0.8723                     & 0.6101                    & 0.6895                       & 0.3034                   & 0.6737                     & 0.2125                    & 0.4100                       & 0.2365                   & 0.4910                     & 0.2026                    & 0.3003                      \\
\multicolumn{1}{c|}{Knowl.} & PDRS                                        & 0.6894          & \textbf{0.8789}            & 0.6390           & \textbf{0.7142}              & \textbf{0.3290}          & \textbf{0.6993}            & \textbf{0.2825}           & \textbf{0.4281}              & \textbf{0.2427}          & \textbf{0.5150}            & 0.2110                    & \textbf{0.3172}             \\ \hline
\end{tabular}
}
\caption{Hit Ratio (H) and NDCG (N) @$K$ on the Course, Movie, and Book datasets. Bold figures highlight best performers.}
\vspace{-0.3cm}
\label{tab:results}
\end{table*}

\begin{table}
\setlength{\abovecaptionskip}{-0.5cm}
\parbox{.56\linewidth}{
    \centering
    \footnotesize
    \begin{tabular}{ccc|cc|cc|cc}
    % \toprule[2pt]
    \toprule
    \multicolumn{1}{c}{\multirow{2}{*}{$\mathcal{C}^p$}} & \multicolumn{1}{c}{\multirow{2}{*}{$\mathcal{C}^t$}} & \multicolumn{1}{c|}{\multirow{2}{*}{$\mathcal{C}^i$}}    & \multicolumn{2}{c|}{$Course$}       & \multicolumn{2}{c|}{$Movie$} & \multicolumn{2}{c}{$Book$} \\ 
    & &  & H@10 & \multicolumn{1}{c|}{N@10} & H@10 & \multicolumn{1}{c|}{N@10} & H@10 & \multicolumn{1}{c}{N@10} \\ 
    
    \midrule
     &  &  &                                 0.8192 & 0.6117 & 0.6444 & 0.3796 & 0.4460 & 0.2749  \\
    \checkmark  &             &            & 0.8626 & 0.6944 & 0.6598 & 0.4003 & 0.4579 & 0.2810  \\
                & \checkmark  &            & 0.8556 & 0.6929 & 0.6714 & 0.4026 & 0.4669 & 0.2859  \\ 
                &             & \checkmark & 0.8422 & 0.6539 & 0.6731 & 0.3991 & 0.4684 & 0.2826  \\            
    \checkmark  & \checkmark  &            & \textit{0.8682} & \textit{0.7099} & 0.6705 & 0.4043 & 0.4644 & 0.2809   \\
    \checkmark  &             & \checkmark & 0.8633 & 0.6968 & \textit{0.6940} & \textit{0.4208} & 0.4898 & 0.3053 \\
                & \checkmark  & \checkmark & 0.8639 & 0.6978 & 0.6851 & 0.4134 & \textit{0.4966} & \textit{0.3062}  \\
    \checkmark  & \checkmark  & \checkmark & \textbf{0.8789} & \textbf{0.7142} & \textbf{0.6993} & \textbf{0.4281} & \textbf{0.5150} & \textbf{0.3172}  \\
    \bottomrule
    \end{tabular}
    \caption{Ablation study on PDRS. $\mathcal{C}^p$, $\mathcal{C}^t$ and $\mathcal{C}^i$ represent user prior concepts, user target concepts, and concepts contained by item. Bold figures indicated leading performers; italicized figures, second-best.}
    \label{tab:variable}
}
\hfill
\parbox{.42\linewidth}{
    \centering
    \footnotesize
    \begin{tabular}{c|cc|cc}
    % \hline
    \toprule
                 & \multicolumn{2}{c|}{\textit{User Cold Start}}                     & \multicolumn{2}{c}{\textit{Item Cold Start}} \\
     & \multicolumn{1}{c}{H@10} & \multicolumn{1}{c|}{N@10} & H@10& N@10      \\ 
    %  \hline
    \midrule
    ItemPop    & 0.7013 & 0.5010     & -  & -    \\
    SVD        & 0.347  & 0.2259     & 0.0661 & 0.0234 \\
    NeuMF      & 0.7097 & 0.4314     & 0.0036 & 0.0012 \\
    BPR        & 0.3787 & 0.2690     & 0.0915 & 0.0414 \\ 
    % \hline
    \midrule
    PDRS                         & \textbf{0.8337} & \textbf{0.6005}    & \textbf{0.1989} & \textbf{0.1110}  \\
    PDRS (w/o $\mathcal{C}_{u/v}$)  & 0.6381 & 0.3977    & 0.0729 & 0.0271     \\ 
    % \hline
    \bottomrule
    \end{tabular}
    \caption{Cold start evaluation on Course recommendation.  PDRS uses knowledge from both user side and item side.  PDRS w/o $\mathcal{C}_{u/v}$ denotes the ablation of user/item context in user/item cold start, respectively.}
    \label{tab:cold}
}
\end{table}

Table~\ref{tab:results} shows recommendation performance.
In general, PDRS outperforms the baselines across all three datasets, in terms of both HR and NDCG, demonstrating the effectiveness of prerequisite context. One exception is that PDRS performs worse than ItemKNN on course recommendation when $k = 2$. We believe this is due to the differing levels of sparsity on the item side: there are $\frac{89K}{4.3K}=20.7$ records per item for courses, but only $\frac{409K}{70K}=5.8$ records per item for books. This makes it easier to find similar items in the former, but more difficult in the latter. This gain evaporates when $k$ increases to 10, as ItemKNN only recommends accurately for users who choose courses similar to previous ones; however we see that users do manifest individualized learning pathways in courses, which ItemKNN does not generalize to. 
GCMC and DeepFM perform well, outperforming other baselines, but they improve dramatically with addition of encoded prerequisite information from KEM, validating that prerequisites play an important role in recommendation tasks.
The variants utilizing KEM pretraining (GCMC+KEM, DeepFM+KEM, PDRS) all improve over their corresponding base models.
Among these three models, PDRS performs best.

% Let us examine PDRS in more depth. 
To verify which form of side information in PDRS is most responsible for performance gains (i.e., user prior knowledge, user target knowledge, and item concept in Table~\ref{tab:variable}), we conduct an ablation study (Table~\ref{tab:variable}). When no side information is used, the model functions just as matrix factorisation. For all three recommendation scenarios, the results are best when all forms of side information are used, and worst when none are leveraged. The result of introducing a single form of side information shows that user context plays the most important role in course recommendation (+\underline{5.3}\% with $\mathcal{C}^p$, +4.4\% with $\mathcal{C}^t$, and +2.8\% with $\mathcal{C}^i$), whereas item context is more helpful for movie and book recommendation 
% (+2.4\% with $\mathcal{K}^p$, +4.1\% with $\mathcal{K}^t$, and +4.5\% with $\mathcal{K}^i$ for movies; and +2.7\% with $\mathcal{K}^p$, +4.7\% with $\mathcal{K}^t$, and +5.0\% with $\mathcal{K}^i$ for books). 
(+\{2.4, 4.1, \underline{4.4}\}\% for movies and +\{2.7, 4.7, \underline{5.0}\}\% for books on \{$\mathcal{C}^p$,$\mathcal{C}^t$,$\mathcal{C}^i$\}, respectively.
Interestingly, the three datasets exhibit different optimal combinations of two forms of side information. For courses, combining user prior and target knowledge is best, likely because learners do choose necessary bridging courses based on their target course.  In contrast, in movie recommendation, the optimal combination is $\mathcal{C}^p$ and $\mathcal{C}^i$. Watchers may choose based more on their experience with relatable plots and characters.  For book recommendation, $\mathcal{C}^t$ and $\mathcal{C}^i$ is the best combination. We surmise that readers choose from their preferred book category (correlated in our PDRS by target knowledge).  

\subsection{Discussion: Is prerequisite context beneficial for Cold-start Problem?}
User cold-start (also termed as new-user) and item cold-start problems \cite{schein2002methods,lam2008addressing} are major recommendation system concerns. Table~\ref{tab:cold} compares PDRS in the course recommendation against baselines in user and item cold-start scenarios. 

From the table, item cold start is more serious than user cold start, as evidenced by the drastic drop in performance in the former case. Missing item interactions --- in addition to high user sparsity --- makes item and user representations inaccurate. Note that as ItemKNN locates items similar to users' previous interactions and ItemPop uses item popularity, both baseline models do not apply to this item cold start scenario. 
The performance of latent factor based approaches (SVD, BPR) severely drops in such cold start scenarios. NeuMF, being able to learn nonlinear relationships, is aware of more complex information from interaction data, yielding comparable HR@10. 
PDRS significantly outperforms baselines in both cold start scenarios, validating the value added by prerequisite knowledge modeling in recommendation accuracy. By comparing the use of user side information in user cold start problem, we observe the advantage of user information. The same applies to items with item information linking items to users through knowledge linkage.

\subsection{Discussion: Is our induced prerequisite knowledge accurate?}
\label{sec:rq1}
PDRS relies on accurate and comprehensive prerequisite inference from documents. To verify the reliability of our automatically compiled prerequisites, 
we also conduct a quality evaluation of our induced prerequisite graphs. 

Recall that during relation inferring in prerequisite graphs (\S~\ref{sec:data_construct}), we train a logistic regression model to predict the strength of prerequisite edges between concepts.
We validate our PKL score fitted from the use of features, and evaluate using precision, recall and F1. 
We take 80\% as training samples and the remaining 20\% as test.
We compare against two baselines in this knowledge learning prediction task: 1) Hyponym Pattern Method (HPM) \cite{wang2016using}: detects whether prerequisites among noun phrases pairs in sentences fulfill 10 lexico-syntactic patterns (e.g., ``\textit{Python} (NP1), one of the \textit{Programming Language} (NP2)''). 
2) Reference Distance (RD) \cite{liang2015measuring}: where
we modify RD's feature extraction methodology to be able to apply it to our task ({\it cf} \ref{sec:data_construct}) by empirically tuning a symmetric threshold $\theta=0.15$, which is used for an $RD(k_i,k_j)$ of $[-1,-\theta), [-\theta,\theta], (\theta,1]$ denoting a posterior, neutral, or prior prerequisite relationship.
% Recall that RD outputs a real-valued score for prerequisite detection between two knowledge concepts. 
% By empirical tuning, we set a symmetric threshold $\theta=0.15$ to determine whether a prerequisite exists; i.e., an $RD(k_i,k_j)$ of $[-1,-\theta), [-\theta,\theta], (\theta,1]$ denoting a posterior, neutral, or prior prerequisite relationship, respectively. 
% {\bf 3.} Finally, we compare against an ablated form of our KPL component by testing only the \textbf{Asymmetric Distance (AsyD)} module. \\

Table~\ref{tab:KPL} shows the macroscopic comparative results against the Course ground truth. Both AsyD and the final
PKL outperform the other two baselines by large margins ($+15$ F1).  The performance improvement is mostly attributed to domain-specific features (AsyD is much better than RD),
whereas general domain features brought in by the final PKL bring a minor boost.
% The microscopic cases (such as \textit{{'Machine learning' $\rightarrow$ 'Deep Learning'}} is recorded by a high prerequisite linkage score $0.78$) further support the enough reliability of our approach. More detailed micro analysis can be found in Appendix~\ref{appendix:micro} and Appendix~\ref{appendix:cases}.
We zoom in on a typical microscopic case study of the  {\it machine learning} knowledge concept from the Course dataset. Table~\ref{tab:pre-sample} lists knowledge concepts that frequently co-occur with {\it machine learning}. Large, positive PKL scores indicate that \textit{machine learning} is a prerequisite of the knowledge concept.  We see that concepts on the right, high-PKL column have higher probability of being recommended to users who master {\it machine learning}, compared to concepts on the left, low-PKL side. As an example, people taking up a course on {\it machine learning} usually have learned \textit{python}. Prerequisite linking in our other two datasets is less intuitive, but meaningful nonetheless. 
% More details can be found in Appendix~\ref{appendix:micro} and Appendix~\ref{appendix:cases}.

\begin{table}
\setlength{\abovecaptionskip}{-0.3cm}
\parbox{.30\linewidth}{
    \setlength{\abovecaptionskip}{-0.2cm}
    \centering
    \footnotesize
    \begin{tabular}{c|ccc}
    \toprule
    Approach & Precision & Recall & F1 \\ \midrule
    HPM & 66.68 & 16.09 & 25.92 \\
    RD & 59.37 & 47.44 & 52.79 \\
    AsyD (Ours) & \textbf{77.08} & 60.66 & 67.89 \\
    PKL (Ours) & 76.00 & \textbf{62.29} & \textbf{68.47} \\
    \bottomrule
    \end{tabular}
    \caption{Prerequisite extraction performance (\%) on our Course dataset. AsyD is an ablated form of our prerequisite inferring component by testing only the Asymmetric Distance (AsyD) module. Bold figures highlight the best performer.}
    \label{tab:KPL}
}
\hfill
\parbox{.65\linewidth}{
    \footnotesize
    \begin{tabular}{c|cl|cl|cl}
    \hline
    Domain & Knowledge           & \multicolumn{1}{c|}{PKL} & Knowledge                & \multicolumn{1}{c|}{PKL} & Knowledge            & \multicolumn{1}{c}{PKL} \\ \hline
    \multicolumn{1}{l|}{\multirow{3}{*}{\textit{Course}}} & Python            & 0.38                     & Code                   & 0.50                     & Feature Learning   & 0.73                    \\
    & Machine Principle & 0.34                     & Program                & 0.50                     & Deep Learning      & 0.78                    \\
    & Computer Basic    & 0.34                     & Database               & 0.50                     & Algorithm Analysis & 0.73                    \\ \hline

    \multicolumn{1}{l|}{\multirow{3}{*}{\textit{Movie}}} & Voldemort  & 0.38   & Policeman            & 0.56             &  Handsome Boy  & 0.75                    \\
    & Triwizard Tournament & 0.34        &  Battle              & 0.50                     & Europe     & 0.78                    \\
    & Evil Dragon & 0.31         &       Merchant         & 0.50       & Fate of Human & 0.93                    \\ \hline
    
    \multicolumn{1}{l|}{\multirow{3}{*}{\textit{Book }}} & Scientist & 0.39  & Atom Bomb & 0.50  & Space Travel   & 0.91           \\
    & 21$^{th}$ Centuries & 0.35   &  Flash & 0.50      & Ethologist    & 0.80       \\
    & World War & 0.35   &    Earthquake  & 0.50      &  Star Trek     &   0.87      \\ \hline
    
    \end{tabular}
    \caption{Example  Prerequisite Knowledge Linkage (PKL) scores for items $PKL$(\textit{`Machine Learning'}, knowledge) from Course, $PKL$(\textit{`Harry Potter'}, knowledge) from Movie, and $PKL$(\textit{`Alien'}, knowledge) from Book.}
    \label{tab:pre-sample}
}

\end{table}

% \subsection{Discussion: Are our PDRS sensitive to hyperparameter?}
% We look into how PDRS handles both knowledge prerequisite and user/item encodings as the model complexity (in terms of hidden layers) is varied. As shown in Table~\ref{tab:pretrain}, models with or without pretraining both perform best with $L=4$.
% Fewer layers are insufficient to learn the complex relationship between embeddings (especially for PDRS to learn the relation between knowledge embedding from both user and item), whereas larger numbers suggest over-fitting. 

% PDRS with pretraining achieves better performance with more hidden layers, but is worse than the case without pretraining, when number of layers is small (i.e., $L=1,2,3$). It can be seen that using pretraining may yield more accurate user/item and knowledge embeddings as initial values for PDRS. Again, too few layers may be insufficiently rich to model embeddings for recommendation. We also exam the sensitivity of our PDRS to the dimension in \ref{appendix:para}

% Holden: @Min. I rewrite the whole section and need a proof reading.
\section{Related Work}
\textbf{Context-aware recommender systems} handle static metadata as auxiliary information, such as user profiles \cite{xu2018graphcar, lei2016comparative}
and item attributes \cite{chen2019personalized}.
Textual content is a typical form with rich contextual information to facilitate RS accuracy. Some works treat item content as raw features by hidden vectors \cite{zheng2017joint, sun2020dual},
while others select important text pieces, such as item tags \cite{gong2016hashtag, li2016hashtag}, and  semantic clues \cite{wang2020fine}.
However, few works focus on establishing context causality between user and item, where our approach uses prerequisite context.

Proper \textbf{Prerequisite Relation Identification} is thus crucial for both intrinsic prerequisite representation task and our ultimate extrinsic task of recommendation. 
Many works rely on statistical methods to determine prerequisites. 
% such as, Bayesian Network, probabilistic association rule, 
An early study by  \citet{vuong2011method} examined the effect of learning curriculum units in various orders. 
\citet{chen2015discovering} treated prerequisite relations as a Bayesian network, which requires a mapping of courses to fine-grained skill and relevant student performance data. 
\citet{chen2016joint} apply probabilistic association rule mining to infer student knowledge from performance data.  
To make the prerequisite relation identification more feasible, others --- including ourselves --- tap into generic information sources. 
\citet{pan2017prerequisite} utilize a Wikipedia corpus to learn semantic representation of concepts for detecting prerequisites in MOOC. \citet{talukdar2012crowdsourced} study how prerequisites can be inferred between Wikipedia entities. \citet{wang2016using} use Wikipedia articles and categories for Concept Graph Learning that uses observed prerequisite relation to learn unobserved ones.
Although this task has mostly been applied to education \cite{yang2015concept}, our findings emphasize that prerequisites indeed generalize and do not need to be restricted to a particular context.  
\section{Conclusion and Future Work}

To the best of our knowledge, we are the first to explore the use of prerequisites --- an overlooked but crucial form of context --- for recommendation.
We introduce Prerequisite Knowledge Linking (PKL) method to induce a prerequisite graph automatically, through semi-supervised learning over both general and domain-specific features.
We instantiate our formalism in the form of a Prerequisite Driven Recommendation System (PDRS; \S3) embodied as a modern neural architecture, which adopts joint training to optimise the model for the twin objectives of knowledge linking prediction and recommendation.
We demonstrate that prerequisite context is a functional booster to solve cold-start problem, and can benefit recommenders universally through our experiments on our Course, Movie, and Book datasets (\S4).

% Holden: @Min. I rewrite the future work in a short paragraph, which needs double checking.
%%%%%%%%%%%%%%%%%%%%%%%%%%%%%
While our PDRS is a simple instantiation of a prerequisite driven recommendation, its elegance leads to synergistic performance gains. Designing more sophisticated models to leverage captured prerequisite knowledge is open  future work. 
As our prerequisite graph is a structured format of knowledge, future work may seek more complex encoding methods, such as typical translation-based methods (e.g., TransE \cite{bordes2013translating}, TransH \cite{wang2014knowledge}).
%as in general knowledge graphs.
Moreover, our work opens the door for studying user's state of knowledge in dynamic scenarios, such as conversational recommendation and long-term sequential recommendation.

%%%%%%%%%%%%%%%%%%%%%%%%%%%%%%% To do
% - Framewrok重新画. k
% - KPL和PLK等保持一致
% - Intro和Conclusion中提及我们的简单性. k
%   - g和f都用的MLP. k
% - 检查引用有误错误
% - Conclusion中加上contribution
% - 花十分钟过一下appendix
% - 把看起来特别复杂的地方简化
% - user knowledge / item knowledge => context
% - 检查K^i的使用
% - knowledge concept, concept, prerequisite concept, prerequisite knowledge的使用
% - RW recheck

% - Recheck 3.2
% - Recheck 3.3
% - Recheck the defintion of PDR and PDRS
% - check conclusion


% \begin{acks}
% We thank Mr. Miao for insightful discussion. We acknowledge the support of NVIDIA Corporation for their donation of the Titan X GPU that facilitated this research. 
% \end{acks}

\bibliographystyle{ACM-Reference-Format}
\bibliography{bibliography}


\begin{thebibliography}{40}

%%% the Web address.

\ifx \showCODEN    \undefined \def \showCODEN     #1{\unskip}     \fi
\ifx \showDOI      \undefined \def \showDOI       #1{#1}\fi
\ifx \showISBNx    \undefined \def \showISBNx     #1{\unskip}     \fi
\ifx \showISBNxiii \undefined \def \showISBNxiii  #1{\unskip}     \fi
\ifx \showISSN     \undefined \def \showISSN      #1{\unskip}     \fi
\ifx \showLCCN     \undefined \def \showLCCN      #1{\unskip}     \fi
\ifx \shownote     \undefined \def \shownote      #1{#1}          \fi
\ifx \showarticletitle \undefined \def \showarticletitle #1{#1}   \fi
\ifx \showURL      \undefined \def \showURL       {\relax}        \fi
% invisible to TeX
\providecommand\bibfield[2]{#2}
\providecommand\bibinfo[2]{#2}
\providecommand\natexlab[1]{#1}
\providecommand\showeprint[2][]{arXiv:#2}

\bibitem[\protect\citeauthoryear{Agrawal, Golshan, and Papalexakis}{Agrawal
  et~al\mbox{.}}{2016}]%
        {agrawal2016toward}
\bibfield{author}{\bibinfo{person}{Rakesh Agrawal}, \bibinfo{person}{Behzad
  Golshan}, {and} \bibinfo{person}{Evangelos Papalexakis}.}
  \bibinfo{year}{2016}\natexlab{}.
\newblock \showarticletitle{Toward data-driven design of educational courses: a
  feasibility study.}
\newblock \bibinfo{journal}{\emph{Journal of Educational Data Mining}}
  \bibinfo{volume}{8}, \bibinfo{number}{1} (\bibinfo{year}{2016}),
  \bibinfo{pages}{1--21}.
\newblock


\bibitem[\protect\citeauthoryear{Berg, Kipf, and Welling}{Berg
  et~al\mbox{.}}{2017}]%
        {berg2017graph}
\bibfield{author}{\bibinfo{person}{Rianne van~den Berg},
  \bibinfo{person}{Thomas~N Kipf}, {and} \bibinfo{person}{Max Welling}.}
  \bibinfo{year}{2017}\natexlab{}.
\newblock \showarticletitle{Graph convolutional matrix completion}.
\newblock \bibinfo{journal}{\emph{arXiv preprint arXiv:1706.02263}}
  (\bibinfo{year}{2017}).
\newblock


\bibitem[\protect\citeauthoryear{Bordes, Usunier, Garcia-Duran, Weston, and
  Yakhnenko}{Bordes et~al\mbox{.}}{2013}]%
        {bordes2013translating}
\bibfield{author}{\bibinfo{person}{Antoine Bordes}, \bibinfo{person}{Nicolas
  Usunier}, \bibinfo{person}{Alberto Garcia-Duran}, \bibinfo{person}{Jason
  Weston}, {and} \bibinfo{person}{Oksana Yakhnenko}.}
  \bibinfo{year}{2013}\natexlab{}.
\newblock \showarticletitle{Translating embeddings for modeling
  multi-relational data}.
\newblock \bibinfo{journal}{\emph{Advances in neural information processing
  systems}}  \bibinfo{volume}{26} (\bibinfo{year}{2013}).
\newblock


\bibitem[\protect\citeauthoryear{Chen, Chen, Xu, Zhang, Cao, Qin, and Zha}{Chen
  et~al\mbox{.}}{2019}]%
        {chen2019personalized}
\bibfield{author}{\bibinfo{person}{Xu Chen}, \bibinfo{person}{Hanxiong Chen},
  \bibinfo{person}{Hongteng Xu}, \bibinfo{person}{Yongfeng Zhang},
  \bibinfo{person}{Yixin Cao}, \bibinfo{person}{Zheng Qin}, {and}
  \bibinfo{person}{Hongyuan Zha}.} \bibinfo{year}{2019}\natexlab{}.
\newblock \showarticletitle{Personalized fashion recommendation with visual
  explanations based on multimodal attention network: Towards visually
  explainable recommendation}. In \bibinfo{booktitle}{\emph{Proceedings of the
  42nd International ACM SIGIR Conference on Research and Development in
  Information Retrieval}}. \bibinfo{pages}{765--774}.
\newblock


\bibitem[\protect\citeauthoryear{Chen, Gonz{\'a}lez-Brenes, and Tian}{Chen
  et~al\mbox{.}}{2016}]%
        {chen2016joint}
\bibfield{author}{\bibinfo{person}{Yetian Chen}, \bibinfo{person}{Jos{\'e}~P
  Gonz{\'a}lez-Brenes}, {and} \bibinfo{person}{Jin Tian}.}
  \bibinfo{year}{2016}\natexlab{}.
\newblock \showarticletitle{Joint discovery of skill prerequisite graphs and
  student Models.}
\newblock \bibinfo{journal}{\emph{International Educational Data Mining
  Society}} (\bibinfo{year}{2016}).
\newblock


\bibitem[\protect\citeauthoryear{Chen, Wuillemin, and Labat}{Chen
  et~al\mbox{.}}{2015}]%
        {chen2015discovering}
\bibfield{author}{\bibinfo{person}{Yang Chen}, \bibinfo{person}{Pierre-Henr
  Wuillemin}, {and} \bibinfo{person}{Jean-Marc Labat}.}
  \bibinfo{year}{2015}\natexlab{}.
\newblock \showarticletitle{Discovering Prerequisite Structure of Skills
  through Probabilistic Association Rules Mining.}
\newblock \bibinfo{journal}{\emph{International Educational Data Mining
  Society}} (\bibinfo{year}{2015}).
\newblock


\bibitem[\protect\citeauthoryear{Devlin, Chang, Lee, and Toutanova}{Devlin
  et~al\mbox{.}}{2018}]%
        {devlin2018bert}
\bibfield{author}{\bibinfo{person}{Jacob Devlin}, \bibinfo{person}{Ming-Wei
  Chang}, \bibinfo{person}{Kenton Lee}, {and} \bibinfo{person}{Kristina
  Toutanova}.} \bibinfo{year}{2018}\natexlab{}.
\newblock \showarticletitle{Bert: Pre-training of deep bidirectional
  transformers for language understanding}.
\newblock \bibinfo{journal}{\emph{arXiv preprint arXiv:1810.04805}}
  (\bibinfo{year}{2018}).
\newblock


\bibitem[\protect\citeauthoryear{Gong and Zhang}{Gong and Zhang}{2016}]%
        {gong2016hashtag}
\bibfield{author}{\bibinfo{person}{Yuyun Gong} {and} \bibinfo{person}{Qi
  Zhang}.} \bibinfo{year}{2016}\natexlab{}.
\newblock \showarticletitle{Hashtag recommendation using attention-based
  convolutional neural network.}. In \bibinfo{booktitle}{\emph{IJCAI}}.
  \bibinfo{pages}{2782--2788}.
\newblock


\bibitem[\protect\citeauthoryear{Gordon, Zhu, Galstyan, Natarajan, and
  Burns}{Gordon et~al\mbox{.}}{2016}]%
        {gordon2016modeling}
\bibfield{author}{\bibinfo{person}{Jonathan Gordon}, \bibinfo{person}{Linhong
  Zhu}, \bibinfo{person}{Aram Galstyan}, \bibinfo{person}{Prem Natarajan},
  {and} \bibinfo{person}{Gully Burns}.} \bibinfo{year}{2016}\natexlab{}.
\newblock \showarticletitle{Modeling concept dependencies in a scientific
  corpus}. In \bibinfo{booktitle}{\emph{Proceedings of the 54th Annual Meeting
  of the Association for Computational Linguistics (Volume 1: Long Papers)}}.
  \bibinfo{pages}{866--875}.
\newblock


\bibitem[\protect\citeauthoryear{Guo, Tang, Ye, Li, and He}{Guo
  et~al\mbox{.}}{2017}]%
        {guo2017deepfm}
\bibfield{author}{\bibinfo{person}{Huifeng Guo}, \bibinfo{person}{Ruiming
  Tang}, \bibinfo{person}{Yunming Ye}, \bibinfo{person}{Zhenguo Li}, {and}
  \bibinfo{person}{Xiuqiang He}.} \bibinfo{year}{2017}\natexlab{}.
\newblock \showarticletitle{DeepFM: a factorization-machine based neural
  network for CTR prediction}.
\newblock \bibinfo{journal}{\emph{arXiv preprint arXiv:1703.04247}}
  (\bibinfo{year}{2017}).
\newblock


\bibitem[\protect\citeauthoryear{He, Liao, Zhang, Nie, Hu, and Chua}{He
  et~al\mbox{.}}{2017}]%
        {he2017neural}
\bibfield{author}{\bibinfo{person}{Xiangnan He}, \bibinfo{person}{Lizi Liao},
  \bibinfo{person}{Hanwang Zhang}, \bibinfo{person}{Liqiang Nie},
  \bibinfo{person}{Xia Hu}, {and} \bibinfo{person}{Tat-Seng Chua}.}
  \bibinfo{year}{2017}\natexlab{}.
\newblock \showarticletitle{Neural collaborative filtering}. In
  \bibinfo{booktitle}{\emph{Proceedings of the 26th international conference on
  world wide web}}. \bibinfo{pages}{173--182}.
\newblock


\bibitem[\protect\citeauthoryear{Hidasi, Karatzoglou, Baltrunas, and
  Tikk}{Hidasi et~al\mbox{.}}{2015}]%
        {hidasi2015session}
\bibfield{author}{\bibinfo{person}{Bal{\'a}zs Hidasi},
  \bibinfo{person}{Alexandros Karatzoglou}, \bibinfo{person}{Linas Baltrunas},
  {and} \bibinfo{person}{Domonkos Tikk}.} \bibinfo{year}{2015}\natexlab{}.
\newblock \showarticletitle{Session-based recommendations with recurrent neural
  networks}.
\newblock \bibinfo{journal}{\emph{arXiv preprint arXiv:1511.06939}}
  (\bibinfo{year}{2015}).
\newblock


\bibitem[\protect\citeauthoryear{Justeson and Katz}{Justeson and Katz}{1995}]%
        {justeson1995technical}
\bibfield{author}{\bibinfo{person}{John~S Justeson} {and}
  \bibinfo{person}{Slava~M Katz}.} \bibinfo{year}{1995}\natexlab{}.
\newblock \showarticletitle{Technical terminology: some linguistic properties
  and an algorithm for identification in text}.
\newblock \bibinfo{journal}{\emph{Natural language engineering}}
  \bibinfo{volume}{1}, \bibinfo{number}{1} (\bibinfo{year}{1995}),
  \bibinfo{pages}{9--27}.
\newblock


\bibitem[\protect\citeauthoryear{Koren, Bell, and Volinsky}{Koren
  et~al\mbox{.}}{2009}]%
        {koren2009matrix}
\bibfield{author}{\bibinfo{person}{Yehuda Koren}, \bibinfo{person}{Robert
  Bell}, {and} \bibinfo{person}{Chris Volinsky}.}
  \bibinfo{year}{2009}\natexlab{}.
\newblock \showarticletitle{Matrix factorization techniques for recommender
  systems}.
\newblock \bibinfo{journal}{\emph{Computer}} \bibinfo{volume}{42},
  \bibinfo{number}{8} (\bibinfo{year}{2009}), \bibinfo{pages}{30--37}.
\newblock


\bibitem[\protect\citeauthoryear{Lam, Vu, Le, and Duong}{Lam
  et~al\mbox{.}}{2008}]%
        {lam2008addressing}
\bibfield{author}{\bibinfo{person}{Xuan~Nhat Lam}, \bibinfo{person}{Thuc Vu},
  \bibinfo{person}{Trong~Duc Le}, {and} \bibinfo{person}{Anh~Duc Duong}.}
  \bibinfo{year}{2008}\natexlab{}.
\newblock \showarticletitle{Addressing cold-start problem in recommendation
  systems}. In \bibinfo{booktitle}{\emph{Proceedings of the 2nd international
  conference on Ubiquitous information management and communication}}.
  \bibinfo{pages}{208--211}.
\newblock


\bibitem[\protect\citeauthoryear{Laurence and Margolis}{Laurence and
  Margolis}{1999}]%
        {laurence1999concepts}
\bibfield{author}{\bibinfo{person}{Stephen Laurence} {and}
  \bibinfo{person}{Eric Margolis}.} \bibinfo{year}{1999}\natexlab{}.
\newblock \showarticletitle{Concepts and cognitive science}.
\newblock \bibinfo{journal}{\emph{Concepts: core readings}}
  \bibinfo{volume}{3} (\bibinfo{year}{1999}), \bibinfo{pages}{81}.
\newblock


\bibitem[\protect\citeauthoryear{Lei, Liu, Li, Zha, and Li}{Lei
  et~al\mbox{.}}{2016}]%
        {lei2016comparative}
\bibfield{author}{\bibinfo{person}{Chenyi Lei}, \bibinfo{person}{Dong Liu},
  \bibinfo{person}{Weiping Li}, \bibinfo{person}{Zheng-Jun Zha}, {and}
  \bibinfo{person}{Houqiang Li}.} \bibinfo{year}{2016}\natexlab{}.
\newblock \showarticletitle{Comparative deep learning of hybrid representations
  for image recommendations}. In \bibinfo{booktitle}{\emph{Proceedings of the
  IEEE conference on computer vision and pattern recognition}}.
  \bibinfo{pages}{2545--2553}.
\newblock


\bibitem[\protect\citeauthoryear{Lei, Zhang, He, Miao, Wang, Chen, and
  Chua}{Lei et~al\mbox{.}}{2020}]%
        {lei2020interactive}
\bibfield{author}{\bibinfo{person}{Wenqiang Lei}, \bibinfo{person}{Gangyi
  Zhang}, \bibinfo{person}{Xiangnan He}, \bibinfo{person}{Yisong Miao},
  \bibinfo{person}{Xiang Wang}, \bibinfo{person}{Liang Chen}, {and}
  \bibinfo{person}{Tat-Seng Chua}.} \bibinfo{year}{2020}\natexlab{}.
\newblock \showarticletitle{Interactive path reasoning on graph for
  conversational recommendation}. In \bibinfo{booktitle}{\emph{Proceedings of
  the 26th ACM SIGKDD International Conference on Knowledge Discovery \& Data
  Mining}}. \bibinfo{pages}{2073--2083}.
\newblock


\bibitem[\protect\citeauthoryear{Li, Liu, Jiang, and Zhang}{Li
  et~al\mbox{.}}{2016}]%
        {li2016hashtag}
\bibfield{author}{\bibinfo{person}{Yang Li}, \bibinfo{person}{Ting Liu},
  \bibinfo{person}{Jing Jiang}, {and} \bibinfo{person}{Liang Zhang}.}
  \bibinfo{year}{2016}\natexlab{}.
\newblock \showarticletitle{Hashtag recommendation with topical attention-based
  LSTM}. In \bibinfo{booktitle}{\emph{Proceedings of COLING 2016, the 26th
  International Conference on Computational Linguistics: Technical Papers}}.
  \bibinfo{pages}{3019--3029}.
\newblock


\bibitem[\protect\citeauthoryear{Liang, Wu, Huang, and Giles}{Liang
  et~al\mbox{.}}{2015}]%
        {liang2015measuring}
\bibfield{author}{\bibinfo{person}{Chen Liang}, \bibinfo{person}{Zhaohui Wu},
  \bibinfo{person}{Wenyi Huang}, {and} \bibinfo{person}{C~Lee Giles}.}
  \bibinfo{year}{2015}\natexlab{}.
\newblock \showarticletitle{Measuring prerequisite relations among concepts}.
  In \bibinfo{booktitle}{\emph{Proceedings of the 2015 conference on empirical
  methods in natural language processing}}. \bibinfo{pages}{1668--1674}.
\newblock


\bibitem[\protect\citeauthoryear{Liang, Ye, Wu, Pursel, and Giles}{Liang
  et~al\mbox{.}}{2017}]%
        {liang2017recovering}
\bibfield{author}{\bibinfo{person}{Chen Liang}, \bibinfo{person}{Jianbo Ye},
  \bibinfo{person}{Zhaohui Wu}, \bibinfo{person}{Bart Pursel}, {and}
  \bibinfo{person}{C Giles}.} \bibinfo{year}{2017}\natexlab{}.
\newblock \showarticletitle{Recovering concept prerequisite relations from
  university course dependencies}. In \bibinfo{booktitle}{\emph{Proceedings of
  the AAAI Conference on Artificial Intelligence}}, Vol.~\bibinfo{volume}{31}.
\newblock


\bibitem[\protect\citeauthoryear{Livne, Unger, Shapira, and Rokach}{Livne
  et~al\mbox{.}}{2019}]%
        {livne2019deep}
\bibfield{author}{\bibinfo{person}{Amit Livne}, \bibinfo{person}{Moshe Unger},
  \bibinfo{person}{Bracha Shapira}, {and} \bibinfo{person}{Lior Rokach}.}
  \bibinfo{year}{2019}\natexlab{}.
\newblock \showarticletitle{Deep context-aware recommender system utilizing
  sequential latent context}.
\newblock \bibinfo{journal}{\emph{arXiv preprint arXiv:1909.03999}}
  (\bibinfo{year}{2019}).
\newblock


\bibitem[\protect\citeauthoryear{Mihalcea and Tarau}{Mihalcea and
  Tarau}{2004}]%
        {mihalcea2004textrank}
\bibfield{author}{\bibinfo{person}{Rada Mihalcea} {and} \bibinfo{person}{Paul
  Tarau}.} \bibinfo{year}{2004}\natexlab{}.
\newblock \showarticletitle{Textrank: Bringing order into text}. In
  \bibinfo{booktitle}{\emph{Proceedings of the 2004 conference on empirical
  methods in natural language processing}}. \bibinfo{pages}{404--411}.
\newblock


\bibitem[\protect\citeauthoryear{Ohland, Yuhasz, and Sill}{Ohland
  et~al\mbox{.}}{2004}]%
        {ohland2004identifying}
\bibfield{author}{\bibinfo{person}{Matthew~W Ohland}, \bibinfo{person}{Amy~G
  Yuhasz}, {and} \bibinfo{person}{Benjamin~L Sill}.}
  \bibinfo{year}{2004}\natexlab{}.
\newblock \showarticletitle{Identifying and removing a calculus prerequisite as
  a bottleneck in Clemson's General Engineering Curriculum}.
\newblock \bibinfo{journal}{\emph{Journal of Engineering Education}}
  \bibinfo{volume}{93}, \bibinfo{number}{3} (\bibinfo{year}{2004}),
  \bibinfo{pages}{253--257}.
\newblock


\bibitem[\protect\citeauthoryear{Pan, Li, Li, and Tang}{Pan
  et~al\mbox{.}}{2017a}]%
        {pan2017prerequisite}
\bibfield{author}{\bibinfo{person}{Liangming Pan}, \bibinfo{person}{Chengjiang
  Li}, \bibinfo{person}{Juanzi Li}, {and} \bibinfo{person}{Jie Tang}.}
  \bibinfo{year}{2017}\natexlab{a}.
\newblock \showarticletitle{Prerequisite relation learning for concepts in
  moocs}. In \bibinfo{booktitle}{\emph{Proceedings of the 55th Annual Meeting
  of the Association for Computational Linguistics (Volume 1: Long Papers)}}.
  \bibinfo{pages}{1447--1456}.
\newblock


\bibitem[\protect\citeauthoryear{Pan, Wang, Li, Li, and Tang}{Pan
  et~al\mbox{.}}{2017b}]%
        {pan2017course}
\bibfield{author}{\bibinfo{person}{Liangming Pan}, \bibinfo{person}{Xiaochen
  Wang}, \bibinfo{person}{Chengjiang Li}, \bibinfo{person}{Juanzi Li}, {and}
  \bibinfo{person}{Jie Tang}.} \bibinfo{year}{2017}\natexlab{b}.
\newblock \showarticletitle{Course concept extraction in moocs via
  embedding-based graph propagation}. In \bibinfo{booktitle}{\emph{Proceedings
  of the Eighth International Joint Conference on Natural Language Processing
  (Volume 1: Long Papers)}}. \bibinfo{pages}{875--884}.
\newblock


\bibitem[\protect\citeauthoryear{Rendle, Freudenthaler, Gantner, and
  Schmidt-Thieme}{Rendle et~al\mbox{.}}{2012}]%
        {rendle2012bpr}
\bibfield{author}{\bibinfo{person}{Steffen Rendle}, \bibinfo{person}{Christoph
  Freudenthaler}, \bibinfo{person}{Zeno Gantner}, {and} \bibinfo{person}{Lars
  Schmidt-Thieme}.} \bibinfo{year}{2012}\natexlab{}.
\newblock \showarticletitle{BPR: Bayesian personalized ranking from implicit
  feedback}.
\newblock \bibinfo{journal}{\emph{arXiv preprint arXiv:1205.2618}}
  (\bibinfo{year}{2012}).
\newblock


\bibitem[\protect\citeauthoryear{Sarwar, Karypis, Konstan, and Riedl}{Sarwar
  et~al\mbox{.}}{2001}]%
        {sarwar2001item}
\bibfield{author}{\bibinfo{person}{Badrul Sarwar}, \bibinfo{person}{George
  Karypis}, \bibinfo{person}{Joseph Konstan}, {and} \bibinfo{person}{John
  Riedl}.} \bibinfo{year}{2001}\natexlab{}.
\newblock \showarticletitle{Item-based collaborative filtering recommendation
  algorithms}. In \bibinfo{booktitle}{\emph{Proceedings of the 10th
  international conference on World Wide Web}}. \bibinfo{pages}{285--295}.
\newblock


\bibitem[\protect\citeauthoryear{Schein, Popescul, Ungar, and Pennock}{Schein
  et~al\mbox{.}}{2002}]%
        {schein2002methods}
\bibfield{author}{\bibinfo{person}{Andrew~I Schein},
  \bibinfo{person}{Alexandrin Popescul}, \bibinfo{person}{Lyle~H Ungar}, {and}
  \bibinfo{person}{David~M Pennock}.} \bibinfo{year}{2002}\natexlab{}.
\newblock \showarticletitle{Methods and metrics for cold-start
  recommendations}. In \bibinfo{booktitle}{\emph{Proceedings of the 25th annual
  international ACM SIGIR conference on Research and development in information
  retrieval}}. \bibinfo{pages}{253--260}.
\newblock


\bibitem[\protect\citeauthoryear{Shani and Gunawardana}{Shani and
  Gunawardana}{2011}]%
        {Shani2011evaluating}
\bibfield{author}{\bibinfo{person}{Guy Shani} {and} \bibinfo{person}{Asela
  Gunawardana}.} \bibinfo{year}{2011}\natexlab{}.
\newblock \showarticletitle{Evaluating recommendation systems}.
\newblock In \bibinfo{booktitle}{\emph{Recommender systems handbook}}.
  \bibinfo{publisher}{Springer}, \bibinfo{pages}{257--297}.
\newblock


\bibitem[\protect\citeauthoryear{Sun, Wu, Zhang, Fu, Hong, and Wang}{Sun
  et~al\mbox{.}}{2020}]%
        {sun2020dual}
\bibfield{author}{\bibinfo{person}{Peijie Sun}, \bibinfo{person}{Le Wu},
  \bibinfo{person}{Kun Zhang}, \bibinfo{person}{Yanjie Fu},
  \bibinfo{person}{Richang Hong}, {and} \bibinfo{person}{Meng Wang}.}
  \bibinfo{year}{2020}\natexlab{}.
\newblock \showarticletitle{Dual learning for explainable recommendation:
  Towards unifying user preference prediction and review generation}. In
  \bibinfo{booktitle}{\emph{Proceedings of The Web Conference 2020}}.
  \bibinfo{pages}{837--847}.
\newblock


\bibitem[\protect\citeauthoryear{Sun, Guo, Yang, Fang, Guo, Zhang, and
  Burke}{Sun et~al\mbox{.}}{2019}]%
        {sun2019research}
\bibfield{author}{\bibinfo{person}{Zhu Sun}, \bibinfo{person}{Qing Guo},
  \bibinfo{person}{Jie Yang}, \bibinfo{person}{Hui Fang},
  \bibinfo{person}{Guibing Guo}, \bibinfo{person}{Jie Zhang}, {and}
  \bibinfo{person}{Robin Burke}.} \bibinfo{year}{2019}\natexlab{}.
\newblock \showarticletitle{Research commentary on recommendations with side
  information: A survey and research directions}.
\newblock \bibinfo{journal}{\emph{Electronic Commerce Research and
  Applications}}  \bibinfo{volume}{37} (\bibinfo{year}{2019}),
  \bibinfo{pages}{100879}.
\newblock


\bibitem[\protect\citeauthoryear{Talukdar and Cohen}{Talukdar and
  Cohen}{2012}]%
        {talukdar2012crowdsourced}
\bibfield{author}{\bibinfo{person}{Partha Talukdar} {and}
  \bibinfo{person}{William Cohen}.} \bibinfo{year}{2012}\natexlab{}.
\newblock \showarticletitle{Crowdsourced comprehension: predicting prerequisite
  structure in wikipedia}. In \bibinfo{booktitle}{\emph{Proceedings of the
  Seventh Workshop on Building Educational Applications Using NLP}}.
  \bibinfo{pages}{307--315}.
\newblock


\bibitem[\protect\citeauthoryear{Vuong, Nixon, and Towle}{Vuong
  et~al\mbox{.}}{2011}]%
        {vuong2011method}
\bibfield{author}{\bibinfo{person}{Annalies Vuong}, \bibinfo{person}{Tristan
  Nixon}, {and} \bibinfo{person}{Brendon Towle}.}
  \bibinfo{year}{2011}\natexlab{}.
\newblock \showarticletitle{A method for finding prerequisites within a
  curriculum.}. In \bibinfo{booktitle}{\emph{EDM}}. \bibinfo{pages}{211--216}.
\newblock


\bibitem[\protect\citeauthoryear{Wang, Wu, Liu, and Xie}{Wang
  et~al\mbox{.}}{2020}]%
        {wang2020fine}
\bibfield{author}{\bibinfo{person}{Heyuan Wang}, \bibinfo{person}{Fangzhao Wu},
  \bibinfo{person}{Zheng Liu}, {and} \bibinfo{person}{Xing Xie}.}
  \bibinfo{year}{2020}\natexlab{}.
\newblock \showarticletitle{Fine-grained interest matching for neural news
  recommendation}. In \bibinfo{booktitle}{\emph{Proceedings of the 58th annual
  meeting of the association for computational linguistics}}.
  \bibinfo{pages}{836--845}.
\newblock


\bibitem[\protect\citeauthoryear{Wang, Ororbia, Wu, Williams, Liang, Pursel,
  and Giles}{Wang et~al\mbox{.}}{2016}]%
        {wang2016using}
\bibfield{author}{\bibinfo{person}{Shuting Wang}, \bibinfo{person}{Alexander
  Ororbia}, \bibinfo{person}{Zhaohui Wu}, \bibinfo{person}{Kyle Williams},
  \bibinfo{person}{Chen Liang}, \bibinfo{person}{Bart Pursel}, {and}
  \bibinfo{person}{C~Lee Giles}.} \bibinfo{year}{2016}\natexlab{}.
\newblock \showarticletitle{Using prerequisites to extract concept maps
  fromtextbooks}. In \bibinfo{booktitle}{\emph{Proceedings of the 25th acm
  international on conference on information and knowledge management}}.
  \bibinfo{pages}{317--326}.
\newblock


\bibitem[\protect\citeauthoryear{Wang, Zhang, Feng, and Chen}{Wang
  et~al\mbox{.}}{2014}]%
        {wang2014knowledge}
\bibfield{author}{\bibinfo{person}{Zhen Wang}, \bibinfo{person}{Jianwen Zhang},
  \bibinfo{person}{Jianlin Feng}, {and} \bibinfo{person}{Zheng Chen}.}
  \bibinfo{year}{2014}\natexlab{}.
\newblock \showarticletitle{Knowledge graph embedding by translating on
  hyperplanes}. In \bibinfo{booktitle}{\emph{Proceedings of the AAAI conference
  on artificial intelligence}}, Vol.~\bibinfo{volume}{28}.
\newblock


\bibitem[\protect\citeauthoryear{Xu, Shen, Liu, and Shen}{Xu
  et~al\mbox{.}}{2018}]%
        {xu2018graphcar}
\bibfield{author}{\bibinfo{person}{Qidi Xu}, \bibinfo{person}{Fumin Shen},
  \bibinfo{person}{Li Liu}, {and} \bibinfo{person}{Heng~Tao Shen}.}
  \bibinfo{year}{2018}\natexlab{}.
\newblock \showarticletitle{Graphcar: Content-aware multimedia recommendation
  with graph autoencoder}. In \bibinfo{booktitle}{\emph{The 41st International
  ACM SIGIR Conference on Research \& Development in Information Retrieval}}.
  \bibinfo{pages}{981--984}.
\newblock


\bibitem[\protect\citeauthoryear{Yang, Liu, Carbonell, and Ma}{Yang
  et~al\mbox{.}}{2015}]%
        {yang2015concept}
\bibfield{author}{\bibinfo{person}{Yiming Yang}, \bibinfo{person}{Hanxiao Liu},
  \bibinfo{person}{Jaime Carbonell}, {and} \bibinfo{person}{Wanli Ma}.}
  \bibinfo{year}{2015}\natexlab{}.
\newblock \showarticletitle{Concept graph learning from educational data}. In
  \bibinfo{booktitle}{\emph{Proceedings of the Eighth ACM International
  Conference on Web Search and Data Mining}}. \bibinfo{pages}{159--168}.
\newblock


\bibitem[\protect\citeauthoryear{Zheng, Noroozi, and Yu}{Zheng
  et~al\mbox{.}}{2017}]%
        {zheng2017joint}
\bibfield{author}{\bibinfo{person}{Lei Zheng}, \bibinfo{person}{Vahid Noroozi},
  {and} \bibinfo{person}{Philip~S Yu}.} \bibinfo{year}{2017}\natexlab{}.
\newblock \showarticletitle{Joint deep modeling of users and items using
  reviews for recommendation}. In \bibinfo{booktitle}{\emph{Proceedings of the
  tenth ACM international conference on web search and data mining}}.
  \bibinfo{pages}{425--434}.
\newblock


\end{thebibliography}

\appendix
\begin{appendices}
\clearpage
\section{Dataset Statistics}
\label{appendix:dataset}
\begin{table}[h!]
\setlength{\abovecaptionskip}{-0.2cm}
\setlength{\belowcaptionskip}{-0cm}
% \footnotesize
\resizebox{0.48\textwidth}{!}{%
\begin{tabular}{lrrrrr}
\hline
Dataset & \# Users & \# Items & \# U--I & \# Concepts & \# C--C Pairs    \\ \hline
Course     & 34,235 & 4,322  & 89,565        & 5,131       & 227,515    \\
Movie & 608    & 7,354  & 49,738        & 12,503      & 8,402,493  \\
Book      & 92,971 & 70,695 & 409,616       & 58,492      & 15,889,953 \\ 
\hline
\end{tabular}
}
\caption{Statistics of the datasets used in our experiments. U--I: user--item interactions; C--C pairs: concept-concept pairs. }
\label{table:dataset}
\end{table}

Detailed dataset statistics are shown in Table~\ref{table:dataset}. Our Course dataset is sparser per user, validating observations that educational activities are hard to recommend without any side information. 
We follow the commonly-used annotation method \cite{liang2017recovering,pan2017prerequisite,gordon2016modeling} in inferring prerequisite relations: manually labeling a small number of concept pairs to train a logistic regression to label the rest. We annotated 300 pairs $(c_i,c_j)$ for each task with $+1$ if $c_i$ is $c_j$'s prerequisite, $-1$ if $c_j$ is $c_i$'s prerequisite, or 0 if no obvious dependency relation exists.  Of the 300 labeled samples $80\%=240$ are used for model training, and the remaining $20\%=60$ are held out for testing. 
The main purpose of this process is to \textit{de-noise} the PKL scores by considering three indicators, discussed in \S~\ref{sec:rq1}.

\section{Detailed Microscopic Case Study of Generated Prerequisite Graphs}
\label{appendix:micro}

As shown in Table~\ref{tab:pre-sample}, prerequisite linking in our other two datasets feels less intuitive, but is meaningful nonetheless. Taking selected concepts from our Movie dataset for the film \textit{'Harry Potter'} as an example, we see that the film is learned as a prerequisite for the knowledge concept \textit{'Fate of Human'}.  This means that the source IMDB and TMDB documents mentioning \textit{'Harry Potter'} mention {\it 'Fate of Human'} but not vice versa, leading to a large PKL score, hence a prerequisite.  Casual inspection of the general Web confirms that {\it Harry Potter} is a foil for subsequent discussions about fate and free will, which makes sense.  While the prerequisite does not constitute a strictly sequential viewing order among {\it 'Harry Potter'} and other movies featuring fate and free will as a central theme, these soft constraints do model inspiration; in Books, people who have read \textit{'Alien'}-themed books may go on to choose similarly themed \textit{'Space Travel'} books, including \textit{'Star Trek'} novels.

We also find that prerequisite graphs have different structural properties. Course prerequisite graphs contain the richest dependency relations (i.e., longer average path and larger average node degree), while graphs for the Movie and Book datasets are sparser and shallower.
Our casual observation of the graphs also reveal that longer prerequisite paths correlate with more specific terms (e.g., {\it 'Voldemort'}), while the shorter paths refer to more general knowledge ({\it 'Love'}).
Both studies show consistency, in that movies and books have fewer dependencies on required prerequisites, as compared to the formal knowledge acquired in courses, but that such soft constraints still aid recommendation accuracy.

\section{The Role of Prerequisites in Recommendation}
\label{appendix:cases}
\subsection{What embedding size is most suited for prerequisite representation?} Here, we use Root Mean Squared Error (RMSE) and R Squared (R2) to measure the fidelity of the KEM prediction; that is, after encoding.  In the figure, KEM (Knowledge Encoding Module) refers to the results from knowledge encoding module alone (ablating recommendation), RM (Recommendation Module) refers to the training of recommendation target without constraints from KEM, and PDRS applies both sets of constraints.

\begin{figure}[h!]
\setlength{\belowcaptionskip}{-0.2cm}
\centering
\includegraphics[width=0.7\linewidth]{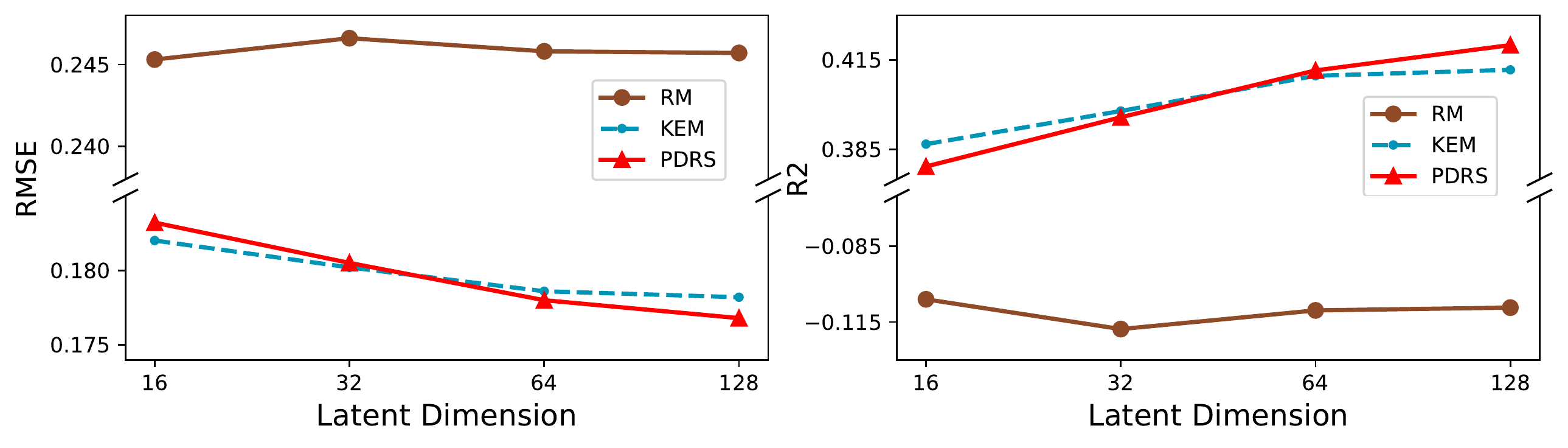}
\caption{Effect of varying the dimensionality of the knowledge concept encoding (x-axes) on downstream recommendation performance (y-axes, as measured in (l) RMSE (lower better) and (r) R2 (higher better).}
\label{fig:klatent_dim}
\end{figure}

Both plots in Figure~\ref{fig:klatent_dim} show a consistent trend: adding more latent dimensions improves knowledge representation and prerequisite capture.
Furthermore, the joint training in PDRS of both $\mathcal{L}_{PKL}$ and $\mathcal{L}_{Rec}$ shows a  slight but consistent benefit.  Both exhibit large improvements compared against optimizing RM alone. 
Seen this way, good prerequisite capture through PKL is its own reward, but has the free side effect of benefitting downstream recommendation as well.

\subsection{Do prerequisites complement user--item interaction?}

Unlike collaborative filtering, prerequisites play a decisive and logical factor in driving users' sequential interactions with items.  While collaborative filtering may find quality recommendations, such recommendations may not account for the immediate needs of the user. 

Take the first row of Table~\ref{tab:micro} as in example of the role of such knowledge.  Here, the left portion of the table gives the input user's state of knowledge and the right portion shows the output recommendation from both the interaction-only BEM and our full PDRS.  
We see this user has handled basic office skills, and thus she may be interested in other related skills such as team work, or soft social skills, which may be recommended by collaborative filtering as is done in BEM.  But perhaps their priority is in building on existing credentials: here, \textit{'Microsoft Word'} can be seen as a more efficient means of  \textit{'text'} entry.

\begin{table*}[h!]
\footnotesize
\abovecaptionskip -0.5mm
\begin{tabular}{|l|ll|}
\hline
\textbf{User Context}  & \textbf{Model} & 
\textbf{First Recommended Item's Knowledge Concepts}  \\ 
\hline
\textbf{PriorK}: 'sort', '\underline{text}', 'ribbon', 'page orientation', 'manipulation' &    \textit{PDRS}  & 'read', 'document', 'page', 'undo', '\underline{microsoft word}'     \\ \cline{2-3}
\textbf{TargetK}: 'file', 'combo box', 'footnote', '\underline{layout}', 'form', 'controller' & \textit{BEM}   & 'team', 'development', 'plan', 'social response', 'application'    \\ 
 
\hline
\textbf{PriorK}: '\underline{muscle}', '\underline{tissues}', 'bone', 'mobile', 'neck' & 
    \textit{PDRS}  & 'autonomic nervous system', 'profile', '\underline{nervous system}', 'area'    \\ \cline{2-3}
\textbf{TargetK}: 'privacies', 'skill', 'certifier', 'participant' & \textit{BEM}   & 'compliance', 'requirement', 'response', 'investigation'     \\ 
  
\hline
\textbf{PriorK}: '\underline{protein}', '\underline{sugar}', 'food product', 'selection' & \textit{PDRS}  & 'bread', '\underline{baking}' \\ \cline{2-3}
\textbf{TargetK}: 'tour', 'participation', 'competition', 'visitor' & \textit{BEM}   & 'preventive act', 'report', 'unit'       \\ 
  
\hline
\textbf{PriorK}: 'conversation', 'spot', 'paper','detection' &  \textit{PDRS}  & '\underline{data analysis}', 'skill', 'sql', 'in demand', 'database'    \\ \cline{2-3}
\textbf{TargetK}: 'database', 'product', 'real world', 'code', '\underline{data feature}', 'universe' & \textit{BEM}   & 'facial', 'neck', 'eye', 'client', 'facial car'    \\ 
\hline
\end{tabular}
\caption{Course examples where PDRS behaves differently. \underline{Underlined concepts} are strongly induced by PKL.}
\label{tab:micro}
\end{table*}

\paragraph{Can prerequisite knowledge be derived from user--item interaction directly?} In short, yes, but much less effectively as we have done in PDRS.
To illustrate the fine-grained level of useful prerequisite knowledge captured (item or concept), we also benchmark PDRS with coarser, obvious prerequisites.   We build another item dependency graph directly from the interaction sequence order in our Course scenario.
We set an aggressive threshold, keeping only the top $\sim$250 item pairs satisfying a minimal number of occurrences and prerequisite strength.  We run these identified, salient prerequisites through BEM but find that performance suffers significantly, leading to a 3.6\% drop in performance. 
This finding direct attributes the power of the fine-grained prerequisite capture, and the effectiveness of PDRS in incorporating general features enhance the signal coming from sequential interaction histories.

\section{Discussion: Is PDRS sensitive to hyperparameter setting?}

We examine how PDRS handles both prerequisite context and user/item encodings as the model complexity (in terms of hidden layers) is varied. As shown in Table~\ref{tab:pretrain}, models with or without pretraining both perform best with $L=4$.
Fewer layers are insufficient to learn the complex relationship between embeddings (especially for PDRS to learn the relation between knowledge embedding from both user and item), whereas larger numbers suggest overfitting. 

PDRS with pretraining achieves better performance with more hidden layers, but is worse than the case without pretraining, when number of layers is small (i.e., $L=1,2,3$). It can be seen that using pretraining may yield more accurate user/item and knowledge embeddings as initial values for PDRS. Again, too few layers may be insufficiently rich to model embeddings for recommendation.

\begin{table}[h!]
\setlength{\abovecaptionskip}{-0.4cm}
\begin{tabular}{|c|c|c|c|c|}
\hline
       & \multicolumn{2}{c|}{\textbf{With Pre-training}} & \multicolumn{2}{c|}{\textbf{Without Pre-training}} \\ \hline
\textbf{\# Layers} & \textbf{HR@10} & \textbf{NDCG@10}  & \textbf{HR@10}  & \textbf{NDCG@10}   \\ \hline
% \textbf{$\Theta$-1} 
1 & 0.7965 & 0.5884    & \textbf{0.8042}  & \textbf{0.5961}    \\ \hline
% \textbf{$\Theta$-2} 
2 & \textbf{0.8543} & 0.6717    & 0.8542 & \textbf{0.6723}              \\ \hline
% \textbf{$\Theta$-3} 
3 & 0.8592 & 0.6937    & \textbf{0.8613} & \textbf{0.6986}              \\ \hline
% \textbf{$\Theta$-4}
4 & \textbf{0.8703}    & \textbf{0.7051}   & 0.8682 & 0.7028              \\ \hline
% \textbf{$\Theta$-5}
5 & \textbf{0.8641}    & \textbf{0.6993}   & 0.8633 & 0.6984              \\ \hline
% \textbf{$\Theta$-6} 
6 & \textbf{0.8640}    & \textbf{0.6989}   & 0.8638 & 0.6932              \\ \hline
\end{tabular}
\caption{Performance of PDRS with varying numbers of layers, with and without pretraining. Bold figures indicate the better strategy w.r.t. each number of layers.}
\label{tab:pretrain}
\end{table}

\begin{figure}[h] 
    \centering
    \setlength{\abovecaptionskip}{-0cm}
    \setlength{\belowcaptionskip}{-0.4cm}
	\subfigtopskip=2pt 
	\subfigbottomskip=2pt 
	\subfigcapskip=-5pt
	\subfigure[HR@10]{
		\includegraphics[width=0.30\linewidth]{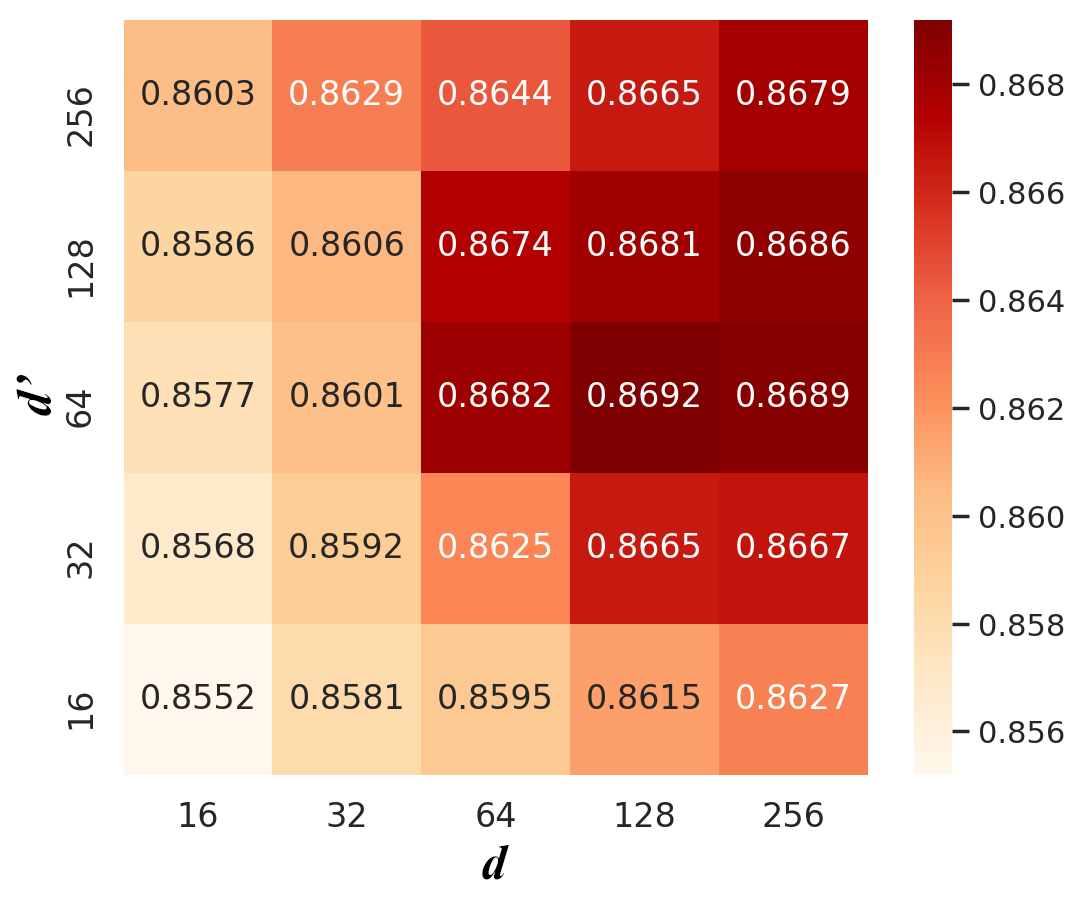}}
% 	\quad 
	\subfigure[NDCG@10]{
		\includegraphics[width=0.30\linewidth]{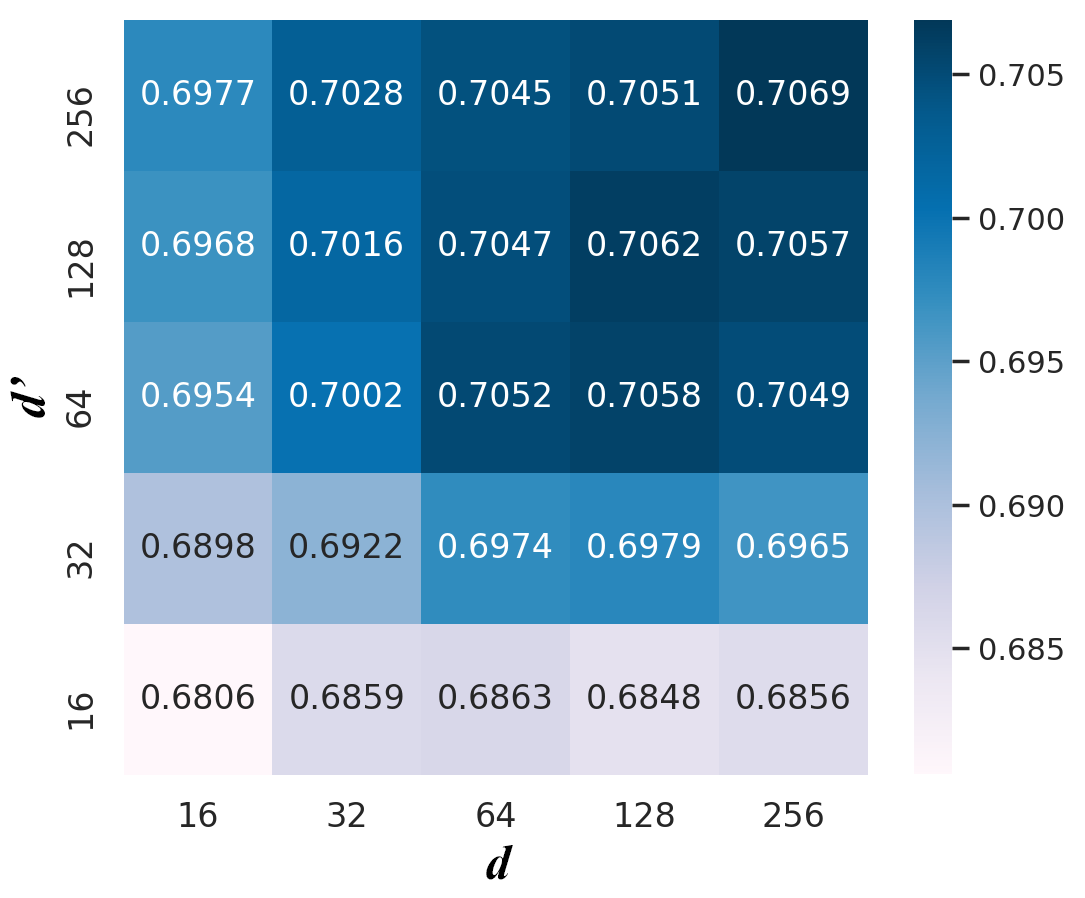}}
	\caption{Top-10 recommendation Hit Ratio and NDCG scores w.r.t. various \# latent factor $d$ in BEM and $d'$ in KEM (\# layer = 4).}
	\label{fig:heatmap}
\end{figure}

We also examine the impact of the combination of the dimension $d$ used for BEM and the dimension $d'$ used for KEM on Course recommendation. As can be seen from the heat map in Figure~\ref{fig:heatmap}, NDCG@10 improves as $d$ and $d'$ increases.  This supports using more factors to store the latent signals and thus improving the model capacity. The best HR@10 is achieved when $d=128, d'=64$, and larger models tend to overfit.

\end{appendices}

\end{document}